\documentclass[fleqn,usenatbib]{mnras}

\usepackage{newtxtext,newtxmath}

\usepackage[T1]{fontenc}

\DeclareRobustCommand{\VAN}[3]{#2}
\let\VANthebibliography\thebibliography
\def\thebibliography{\DeclareRobustCommand{\VAN}[3]{##3}\VANthebibliography}

\DeclareRobustCommand{\DE}[3]{#2}
\let\DEthebibliography\thebibliography
\def\thebibliography{\DeclareRobustCommand{\DE}[3]{##3}\DEthebibliography}

\usepackage{graphicx}	
\usepackage{amsmath}	
\usepackage[version=4]{mhchem}
\usepackage{multirow}
\usepackage{booktabs}

\graphicspath{{Figures/}}

\makeatletter
\newcounter{chemeqn}
\newenvironment{chemequations}{\let\c@equation\c@chemeqn}{}
\makeatother

\newcommand{\hox}{HO$_\text{x}$ }
\newcommand{\nox}{NO$_\text{x}$ }

\newcommand{\pd}[2]{\frac{\partial #1}{\partial #2}}

\title[Stellar activity and tidally locked exoplanets]{3D modelling of the impact of stellar activity on tidally locked terrestrial exoplanets: atmospheric composition and habitability}

\author[Ridgway, R. J.]{R. J. Ridgway$^{1}$\thanks{E-mail: \href{mailto:rr364@exeter.ac.uk}{rr364@exeter.ac.uk}},
M. Zamyatina$^{1}$,
N. J. Mayne$^{1}$,
J. Manners$^{1,2}$,
F. H. Lambert$^{3}$,
M. Braam$^{4,5}$,\newauthor
B. Drummond$^{2}$,
E. H\'{e}brard$^{1}$,
P. I. Palmer$^{4,5}$,
and K. Kohary$^{1}$
\\
$^{1}$Physics and Astronomy, Faculty of Environment, Science and Economy, University of Exeter, Exeter, EX4 4QL, UK\\
$^{2}$Met Office, Fitzroy Road, Exeter, EX1 3PB, UK\\
$^{3}$Mathematics and Statistics, Faculty of Environment, Science and Economy, University of Exeter, Exeter, EX4 4QF, UK\\
$^{4}$School of GeoSciences, The University of Edinburgh, Edinburgh, EH9 3FF, UK\\
$^{5}$Centre for Exoplanet Science, The University of Edinburgh, Edinburgh, EH9 3FD, UK\\
}

\date{Accepted XXX. Received YYY; in original form ZZZ}

\pubyear{2022}

\begin{document}
\label{firstpage}
\pagerange{\pageref{firstpage}--\pageref{lastpage}}
\maketitle

\begin{abstract}
Stellar flares present challenges to the potential habitability of terrestrial planets orbiting M dwarf stars through inducing changes in the atmospheric composition and irradiating the planet's surface in large amounts of ultraviolet light. To examine their impact, we have coupled a general circulation model with a photochemical kinetics scheme to examine the response and changes of an Earth-like atmosphere to stellar flares and coronal mass ejections. We find that stellar flares increase the amount of ozone in the atmosphere by a factor of 20 compared to a quiescent star. We find that coronal mass ejections abiotically generate significant levels of potential bio-signatures such as \ce{N2O}. The changes in atmospheric composition cause a moderate decrease in the amount of ultraviolet light that reaches the planets surface, suggesting that while flares are potentially harmful to life, the changes in the atmosphere due to a stellar flare act to reduce the impact of the next stellar flare.
\end{abstract}




\section{Introduction}
\label{sec:introduction}
The study of potentially habitable terrestrial exoplanets orbiting M dwarfs is likely to play a pivotal role in answering one of the most significant and long--standing questions facing humankind: whether life on Earth is a unique and singular occurrence. This major, overarching, goal encompasses and requires contributions from a wide range of research disciplines. Exoplanet research has an opportunity to make a vital contribution to help unravel this puzzle. 

For the foreseeable future terrestrial planets orbiting M dwarf stars represent our best opportunity of potentially identifying a habitable world beyond the solar system. As we currently have evidence of only one inhabited planet, Earth, we are focusing efforts to identify potentially habitable worlds through the lens of life on Earth. However, although efforts are underway to identify targets amenable to follow--up characterisation as similar to the Earth as possible in terms of host star, orbital parameters etc. (for example, the Terra Hunting Experiment \citet{Thompson2016}\footnote{\url{https://www.terrahunting.org/}}), currently our sample of potentially habitable planets, essentially defined as planets orbiting at a distance from their star such that liquid water could be present on their surface \citep[i.e. in the `habitable zone', or HZ,][]{Kasting1993}, are dominated by those orbiting M dwarfs. Planets orbiting M dwarfs in the HZ must have shorter periods than those orbiting sun--like G dwarfs, due to the lower luminosity of the central star. This shorter period combined with the ubiquity of M dwarfs and the more favourable planet--to--star radius and mass ratios make detection and atmospheric characterisation much more feasible for these planets compared to their G dwarf counterparts. For some of these M dwarf hosted planets it may even be possible to obtain constraints on their atmospheric compositions in the near future, vital for determining potential climates \citep{DeWit2018_Fixed}.

However, there are several difficult challenges to our ability to understand and interpret the climates of any particular target. M dwarfs are cooler, smaller and often much more prone to stellar activity than G dwarfs. The M dwarf HZ is so close to the star that tidal forces are expected to rapidly force the planet into a circular orbit and becoming tidally locked (the same side of the planet always faces the star) \citep{Barnes2017}. Being tidally locked has significant consequences for the planetary climate, primarily a large contrast in the day-night irradiation. This contrast leads to, for example, planetary--scale circulation through a super-rotating equatorial jet \citep{Showman2013}, and large day--night temperature contrasts.


Our understanding, and therefore predictive capability, of the basic climate of terrestrial planets hosted by M dwarfs is rapidly improving. Models of varying complexity have been applied to such planets, starting with the pioneering study of \citet{Joshi1997} and recently with the THAI model intercomparison project \citep[TRAPPIST Habitable Atmospheres Intercomparison,][]{Fauchez2021a,Turbet2021,Sergeev2021,Fauchez2021}. Due to being similar in size to Earth and orbiting their host stars in the HZ, many simulations have focused on two major targets of interest, Proxima Centauri b \citep[ProxCen~b,][]{Anglada-Escude2016} and the TRAPPIST--1 planets \citep{Gillon2017} such as \citet{Turbet2016b} and \citet{Turbet2018}, respectively.


Our understanding of stellar activity of M dwarfs is also increasing. \citet{Gunther2020} performed a study of the first data release from TESS (Transiting Exoplanet Survey Satellite) to look at the population of flaring stars. They found that the majority of flaring stars observed by TESS were M dwarfs. \citet{Hawley2014} and \citet{Davenport2014} found that flares on M dwarfs can occur over a wide range of durations and magnitudes. One of the largest solar flares ever observed, the 1859 Carrington event, was estimated to have released $\approx 5\times10^{32}$\,ergs \citep{Cliver2013}. \citet{Hawley2014} found that for active M dwarfs such as GJ~1243, flares of comparable magnitude can occur approximately once a month. Flares and stellar activity give rise to an increase in the high--energy and short--wavelength emission from the star, alongside releases of energetic particles known as a Coronal Mass Ejection (CME) \citep{Yashiro2006}. These particles are highly energetic, and are capable of inducing changes in the atmospheric composition of terrestrial planets. \citet{Yashiro2006} found that energetic solar flares are almost always accompanied by a CME.

The UV flux received by the surface of Earth has a significant impact on life, and is also believed to play a significant role in the early evolution of organic compounds. On Earth, ever since the Great Oxygenation Event, an event in Earth's history that occurred approximately 2 billion years ago, where the amount of molecular oxygen increased from negligible levels to a concentration similar to modern amounts, the presence of ozone (\ce{O3}) \citep[and potentially organic hazes in the case of the Archean Earth,][]{Arney2016} in the upper atmosphere has acted to regulate the received surface UV flux \citep{Gebauer2017}. Understanding the potential surface UV flux for target M dwarf hosted planets with a similar atmospheric composition is, therefore, an important endeavour, linked to the presence of ozone. 


UV radiation and energetic particles alter the chemistry, and therefore composition, of planetary atmospheres. For the case of modern Earth, most UV radiation is absorbed by ozone at medium--high altitudes, with ozone generated and recycled through an ozone--oxygen cycle commonly called the Chapman cycle \citep{Chapman1930}. Ozone chemistry also depends on the generation of short--lived free radical species termed \hox (\ce{H}, \ce{OH}, and \ce{HO2}) and \nox (\ce{N}, \ce{NO}, and \ce{NO2}), which play an important role in regulating the abundance of ozone. Alongside the impacts of the UV flux the energetic particles emitted from the star ionise the gases in the atmosphere, creating additional \hox and \nox species which contribute to the depletion of atmospheric ozone \citep{Segura2010,Tilley2019}.

The impact of stellar flares and CMEs on terrestrial exoplanets has been addressed in only a small number of studies. Such studies have, however, shown the significant changes they can cause in the chemical processes and composition. For example, \citet{Segura2010} found that according to their results from a 1D photochemical model, for an unmagnetised `Earth-like' planet orbiting an M dwarf star the amount of ozone in the atmosphere was not significantly impacted by a single stellar flare, when only including the increase in electromagnetic radiation. However, they also showed that ozone was initially significantly depleted by the proton flux associated with the stellar flare and CME, before recovering to the original levels of ozone abundance. \citet{Tilley2019} extended on the work of \citet{Segura2010} using the same model but including multiple flares, suggesting that the recovery of the ozone after the period of activity was unlikely. Quite recently, \citet{Louca2022} used a 1D model to examine the impact of stellar activity on a range of atmospheres, from hydrogen (\ce{H2})-dominated to nitrogen (\ce{N2})-dominated, and found potentially permanent changes in the atmospheric composition due to flares. Their \ce{N2} dominated atmosphere simulations showed that flares can cause a gradual increase in the amount of ozone.

The inherently 3D nature of planetary atmospheres are not necessarily captured by a 1D model. Extension from 1D to 3D is expected to give a more plausible description of the planet, especially given that the target planets are expected to be tidally locked, with one hemisphere constantly irradiated, and in the likely absence of a significant magnetic field \citep{Christensen2009}, a hemisphere which is not directly impacted by any stellar activity. 

\citet{Chen2020} performed a 3D study exploring the impact of stellar flares from a range of stellar types (M, K, G) on an Earth-like planet. They found that in the case of K/M dwarfs the planet retained a significantly perturbed atmospheric composition due to flaring (planets around a G dwarf quickly returned to their pre--flare atmospheric composition). This presents questions regarding the interpretation of exoplanetary atmospheres determined using atmospheric retrieval, as nitrous oxide (\ce{N2O}) was found to be significantly enhanced compared to the same planet subject to a non--flaring star's irradiation. On the modern Earth, nitrous oxide's abundance is heavily controlled by biological activity \citep{Syakila2011} (although not entirely, \ce{N2O} can be created abiotically), and is thought to be a bio--signature \citep{DesMarais2002a}. Finding a plausible abiotic source of significant amounts of nitrous oxide would raise doubts about the potential of \ce{N2O} as a bio--signature.

In this work we apply the adapted Met Office Unified Model (UM), which is a 3D General Circulation Model (GCM) to study the impact of stellar activity on an example, potentially habitable M dwarf orbiting planet. We base this study on Proxima Centauri b (hereafter, ProxCen~b), although our conclusions are generally applicable to other M dwarf hosted (e.g., the TRAPPIST planets), potentially habitable, planets. We include treatment of the relevant chemical processes controlling the ozone concentration and distribution, alongside other key species, and also include the impact of energetic particles. The rest of this paper is laid out as follows. In Section~\ref{sec:model_description} we describe the major model components we have added, building on the setup described in \citet{Boutle2017}, for this work, including details of the implementation of the photolysis and stellar protons (Sections~\ref{subsec:photolysis} and \ref{subsec:protons}, respectively). We present our results in Section~\ref{sec:results}, separating the analysis between an initial quiescent phase (Section~\ref{subsec:quiet}) and subsequent, fine temporal resolution simulations including stellar flaring (Section~\ref{subsec:flaring}). We then draw conclusions, and comment on opportunities for future work and development in Sections~\ref{sec:conclusions} and \ref{subsec:future_work}, respectively.

\section{Model Description}
\label{sec:model_description}
\begin{figure*}
    \centering
    \includegraphics[width=\columnwidth]{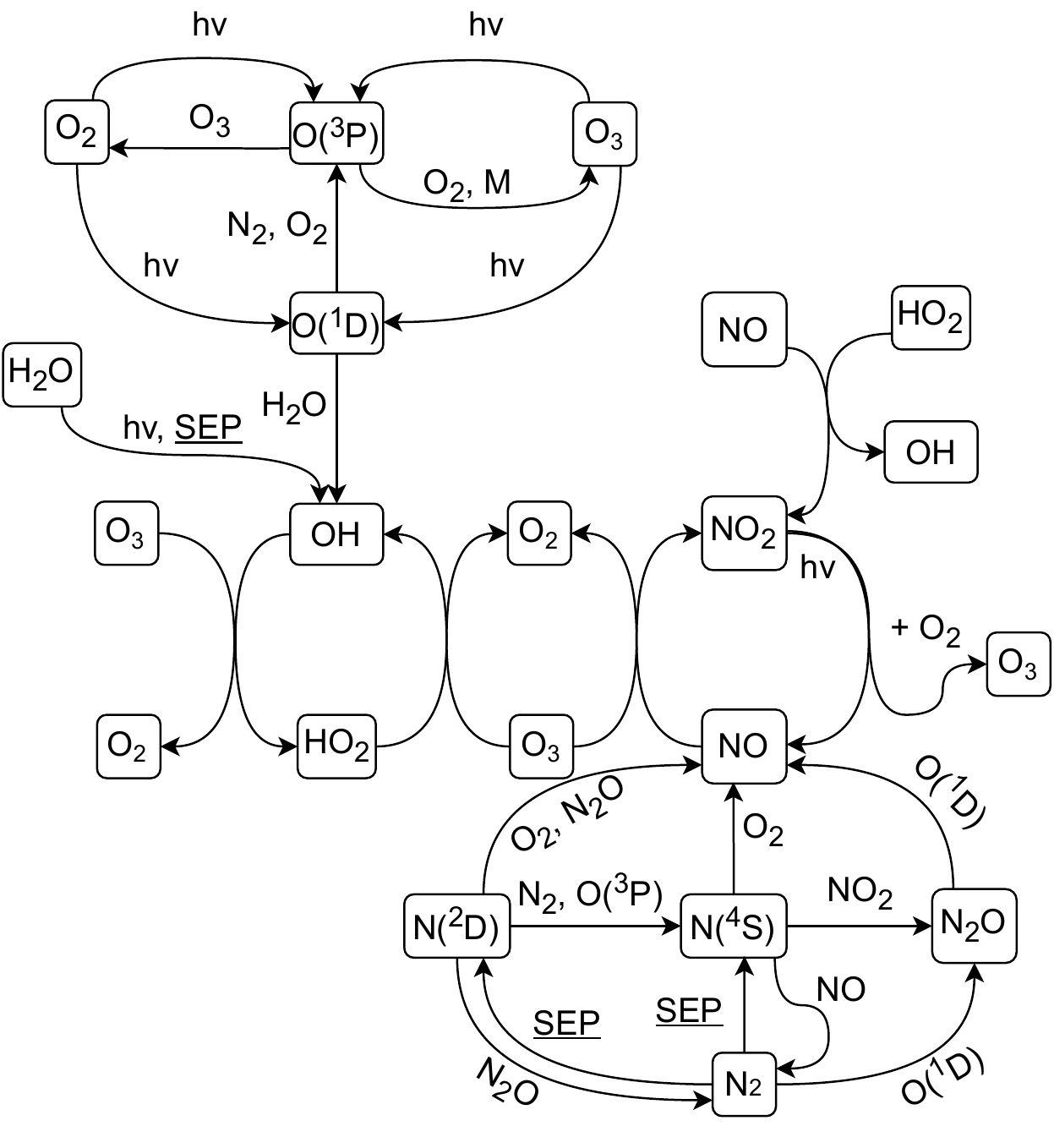}
    \caption{Schematic of the main processes controlling ozone chemistry, and the major species involved, included in our model. hv denotes a photon and describes photolysis reactions. M denotes a termolecular reaction. SEP denotes a reaction caused by stellar energetic particles. For more information including a comprehensive list of all reactions in the chemical network, see Appendix~\ref{appsec:reactions}.}
    \label{Fig:Schematic}
\end{figure*}

In this section we first describe our 3D model framework before detailing the stellar spectrum and  processes affecting the atmospheric chemistry. Figure~\ref{Fig:Schematic} shows a schematic of the main processes and species represented in our simulation that are involved in ozone chemistry. The full list of reactions within our chemical network and the species involved in chemistry are given in Appendix \ref{appsec:reactions}.

\subsection{The Unified Model}
\label{SubSec:um}
The UM is a well verified Earth GCM used seamlessly for both weather and climate prediction \citep{Walters2019}. We have adapted the UM for the study of a wide range of exoplanets, beginning with testing and benchmarking for both terrestrial \citep{Mayne2014b} and hot Jupiter type planets \citep[gas giants in short period orbits,][]{Mayne2014a,Amundsen2014,Amundsen2016}. For the case of tidally locked terrestrial planets orbiting M dwarfs, the UM has been used before to explore the overall climate of an `Earth--like' (by mass, $\sim23$\% \ce{O2}, $\sim0.06$\% \ce{CO2}, $\sim76$\% \ce{N2}, and additional gases such as water vapour) atmospheric composition \citep{Boutle2017}, the impact of the surface on the overall climate \citep{Lewis2018}, the balance of radiation absorption between the surface and atmosphere \citep{Eager2020}, the importance of the treatments of convection to the planetary conditions \citep{Sergeev2021}, and the impacts of dust \citep{Boutle2020}.


The UM has already been used to study the impact of the quiescent spectrum of an M dwarf on the atmospheric ozone distribution of a planet with an initial idealised Earth atmospheric composition \citep{Yates2020}. \citet{Yates2020} used the UM, coupled to the UKCA\footnote{\url{https://www.ukca.ac.uk/wiki/index.php/UKCA}} chemical framework, to model ozone chemistry on a tidally locked planet (based on ProxCen~b). They assumed the planet orbits a quiescent M dwarf, with a pre--industrial Earth-like atmosphere, with atmospheric chemistry consisting of the Chapman cycle and the \hox (in their case defined as \ce{OH}, \ce{HO2}, without any \ce{H}) chemistry. In parallel, we have developed an idealised chemistry framework \citep{Drummond2016}, coupled to the UM, designed to be flexible in terms of both the input chemical network and the level of sophistication ranging from simple equilibrium chemistry \citep{Drummond2018}, to `chemical relaxation' \citep{Drummond2018a,Drummond2018c} and on to full chemical kinetics \citep[][\& Zamyatina et al., submitted]{Drummond2020}. However, this framework has, thus far, been applied only to hot Jupiter planets.

In this work we adapt our chemical kinetics framework \citep{Drummond2020} and radiative transfer scheme \citep[Suite Of Community RAdiative Transfer codes based on][SOCRATES]{Edwards1996} for the study of the Chapman cycle, \hox \& \nox chemistry, alongside including a treatment of the ionisation caused by energetic particles \citep[see][for a technical description of SOCRATES]{Manners2022}. Our implementation, and this study are focused on the stratospheric ozone distribution on an `Earth--like' planet with an initial atmospheric composition similar to modern Earth in a tidally locked orbit of a host M dwarf star, based on ProxCen~b. This work has been performed alongside and in close collaboration with that of \citet{Braam2022} who further adapted the UKCA framework, building on the work of \citet{Yates2020}, to study the impact of lightning on the tropospheric ozone composition of ProxCen~b also using the UM. As we have been using the same GCM and the same planetary configuration, but different chemistry schemes, this has provided a useful environment for testing and developing both of our models. 

Our setup is built upon that presented in \citet{Boutle2017}. We have adopted their planetary parameters  (Table~\ref{Tab:PlanetConstants}), and their `Earth--like' initial atmospheric composition \citep[see Table 2 of][]{Boutle2017}. The orbital parameters are the values measured by \citet{Anglada-Escude2016}. The planet is assumed to be tidally locked. The quiescent stellar constant is calculated using the same inputs and methodology as \citet{Boutle2017}. The planetary radius and surface gravity are those estimated by \citet{Turbet2016b} assuming a mass of 1.4 Earth masses and a density similar to Earth of 5500\,kg m$^{-3}$.The atmosphere is assumed to be \ce{N2-O2} dominated, with \ce{CO2}, and an active water cycle generating water vapour. However, in order to capture the impacts of stellar activity we have updated the treatment of radiative transfer and included atmospheric chemistry and photolysis to respond to a new time-dependent stellar spectrum.




\begin{table}
\caption{he planetary and orbital parameters used in this work. The planet is assumed to be tidally locked.}
\label{Tab:PlanetConstants}
\begin{tabular}{ll}
Planet Constants                               & Proxima Centauri b             \\ \hline  \hline
\multicolumn{1}{l|}{Planet radius (km)}        & 7160                  \\
\multicolumn{1}{l|}{Solar constant (W/m$^2$)}  & 2.07                  \\
\multicolumn{1}{l|}{Rotation rate (radians/s)} & $6.501\times 10^{-6}$ \\
\multicolumn{1}{l|}{Semi-major axis (AU)}      & 0.0485                \\
\multicolumn{1}{l|}{Surface gravity (m/s$^2$)} & 10.9                  \\
\multicolumn{1}{l|}{Eccentricity}              & 0                     \\
\multicolumn{1}{l|}{Obliquity}                 & 0                    
\end{tabular}
\end{table}

\subsection{Radiative transfer}
\label{subsec:radiative}

The SOCRATES configuration files (also known as spectral files) were updated from the files used by \citet{Boutle2017} and have been adapted to account for short-wavelength radiative transfer and the inclusion of photolysis. In this work, we use 16 `shortwave' (stellar radiation) wavelength bands ranging from 0.5\,nm-10\,$\mu$m, and are listed in Table~\ref{Tab:Bands}. Nine `longwave' (thermal emission) bands are used, ranging from 3.3\,$\mu$m--10\,mm. We use the correlated--k technique. We include Rayleigh scattering, and scattering and absorption by liquid and ice clouds. Clouds are described using the PC2 scheme \citep{Wilson2008}, and are coupled to radiative transfer using the MCICA scheme \citep{Pincus2003}. Photolysis is directly calculated alongside radiative heating within SOCRATES, and is described in Section~\ref{subsec:photolysis}. Changes in atmospheric composition due to the coupled chemistry framework are reflected in SOCRATES by changes in radiative heating and photolysis. For a list of all chemical species tracked in our model, and sources for their opacity, refer to Table~\ref{Tab:ChemistrySpecies}. 

\begin{table}
    \caption{The shortwave wavelength bands used by the radiation scheme for this work.}
    \label{Tab:Bands}
    \begin{tabular}{ll}
        Band & Wavelength range (nm) \\ \hline  \hline
        1  & 0.5-75     \\
        2  & 75-100     \\
        3  & 100-125    \\
        4  & 125-150    \\
        5  & 150-175    \\
        6  & 175-200    \\
        7  & 200-225    \\
        8  & 225-250    \\
        9  & 275-300    \\
        10 & 300-320    \\
        11 & 320-505    \\
        12 & 505-690    \\
        13 & 690-1190   \\
        15 & 1190-2380  \\
        16 & 2380-10000
    \end{tabular}
\end{table}

The effects of atmospheric aerosols were ignored. In order to calculate the surface UV environment and transmission spectra (described in Section~\ref{Sec:PlanetHab} and \ref{Sec:PlanetObs} respectively) high resolution spectral files were created to describe their respective wavelength ranges.

\subsection{Stellar Spectrum}
\label{subsec:spectra}
\citet{Boutle2017} and \citet{Yates2020} used a stellar spectrum for ProxCen from BT-Settl (a model of stellar atmospheres) \citep{Rajpurohit2013} assuming an effective temperature $T_{\text{eff}}=3000$\,K, a stellar surface gravity of $g = 1000$\,m s$^2$, and a metallicity of 0.3\,dex. This spectrum includes essentially no UV light below 200\,nm, as the BT-Settl models capture stellar photospheric emission but do not account for chromospheric emission. The study of \citet{Boutle2017} employed a fixed ozone layer, and focused on altitudes below those employed here, meaning the impacts of the missing very short wavelength flux would have been negligible in their study. However, with our focus on the ozone chemistry and higher altitude atmosphere, it is vital we improve on this aspect. Simulations of Earth--like planets over a range of M dwarfs for active and inactive stellar models \citep{Rugheimer2015b} show that models such as BT-Settl (an inactive stellar model) will produce significantly different ozone compositions than more active (models built to describe observations of active M dwarfs) stellar spectra models, and spectra derived from observations of M dwarf stars.

\begin{figure}
    \centering
    \includegraphics[width=\columnwidth]{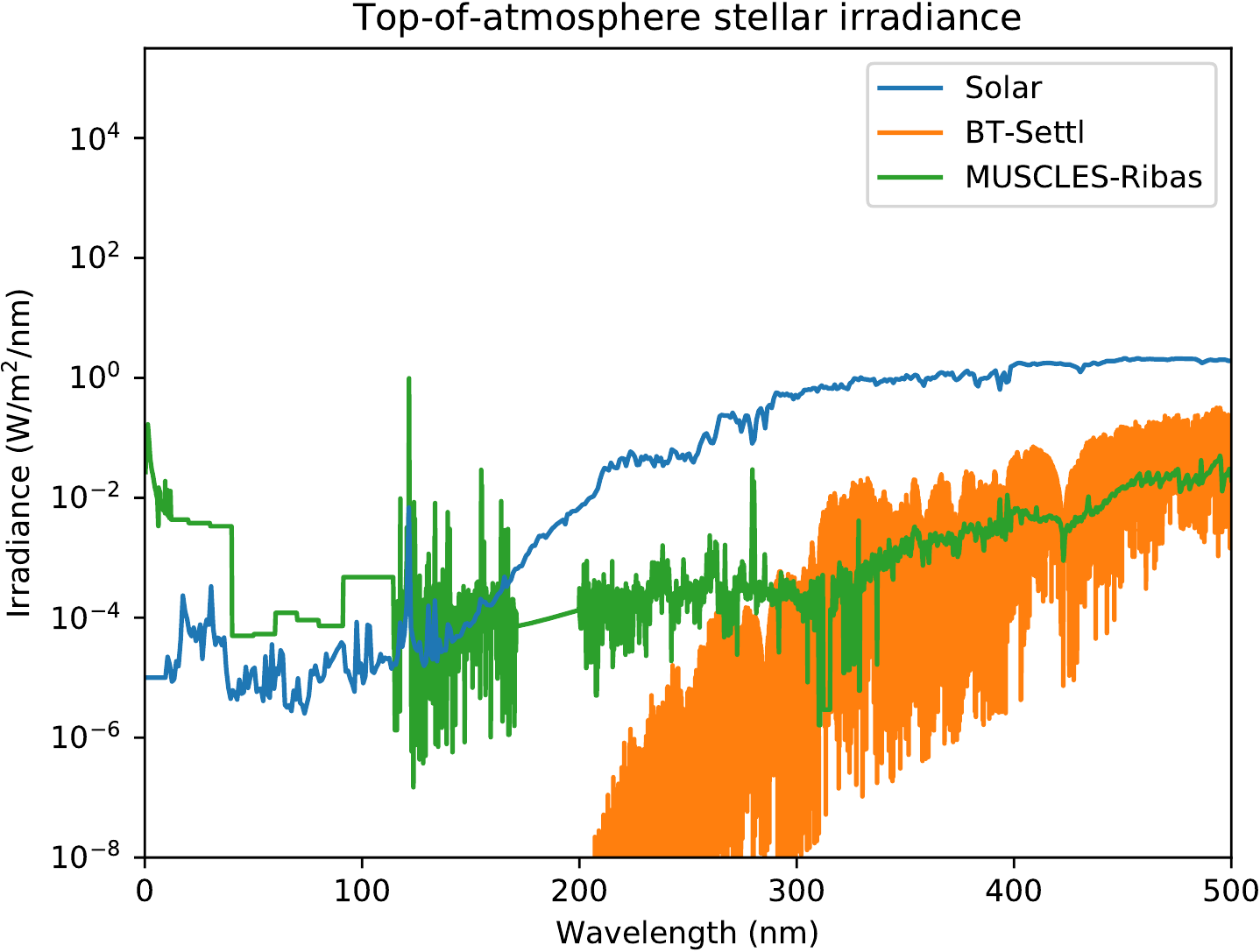}
    \caption{The top-of-atmosphere stellar irradiance for Earth, the BT-Settl spectrum for ProxCen~b, and the combined MUSCLES-Ribas spectrum for ProxCen~b. We note that the irradiance below 300\,nm differs significantly between the two ProxCen~b spectra.}
    \label{Fig:Spectra}
\end{figure}

We have constructed a stellar spectrum from a combination of the MUSCLES survey \citep{France2016,Youngblood2016,P.Loyd2016}\footnote{The \textit{adapt-const-res-sed.fits} version of the spectra} and \citet{Ribas2017} describing ProxCen. This spectrum has significantly higher fluxes in the UV to X-ray than the equivalent BT--Settl model, a significantly different radiation environment. Figure~\ref{Fig:Spectra} illustrates the differences in stellar spectra by showing the top--of--atmosphere stellar irradiance received by the Earth, the BT--Settl model of ProxCen~b, and the spectrum used in this work. The combined spectrum has significantly higher UV radiation than the BT-Settl model below 300\,nm, and in fact has higher levels of extreme UV/X--ray than the Solar spectrum below 120\,nm. The enhancement of UV leads to increased \ce{O2} photolysis (which occurs below 242\,nm). The increased rate of photolysis leads to significantly higher abundances of atomic oxygen, which leads to significantly faster growth in ozone through the three-body reaction
\begin{chemequations}
    \ce{O2 + O(^3P) + M -> O3 + M},
\end{chemequations}
where M denotes a third body. \citet{Braam2022} used the same spectrum as this work to simulate a similar `Earth--like' planet as \citet{Yates2020}, and found that when compared to \citet{Yates2020} (who used the BT-Settl spectrum) they had significantly higher amounts of ozone. This was due to two factors, the change in spectrum to one which has higher UV increases the amount of ozone significantly, and an improved calculation of photolysis rates as compared to the work done by \citet{Yates2020}.

\subsection{Flare Generation}
\label{subsubsec:flare_profile}
\begin{figure}
    \centering
    \includegraphics[width=\columnwidth]{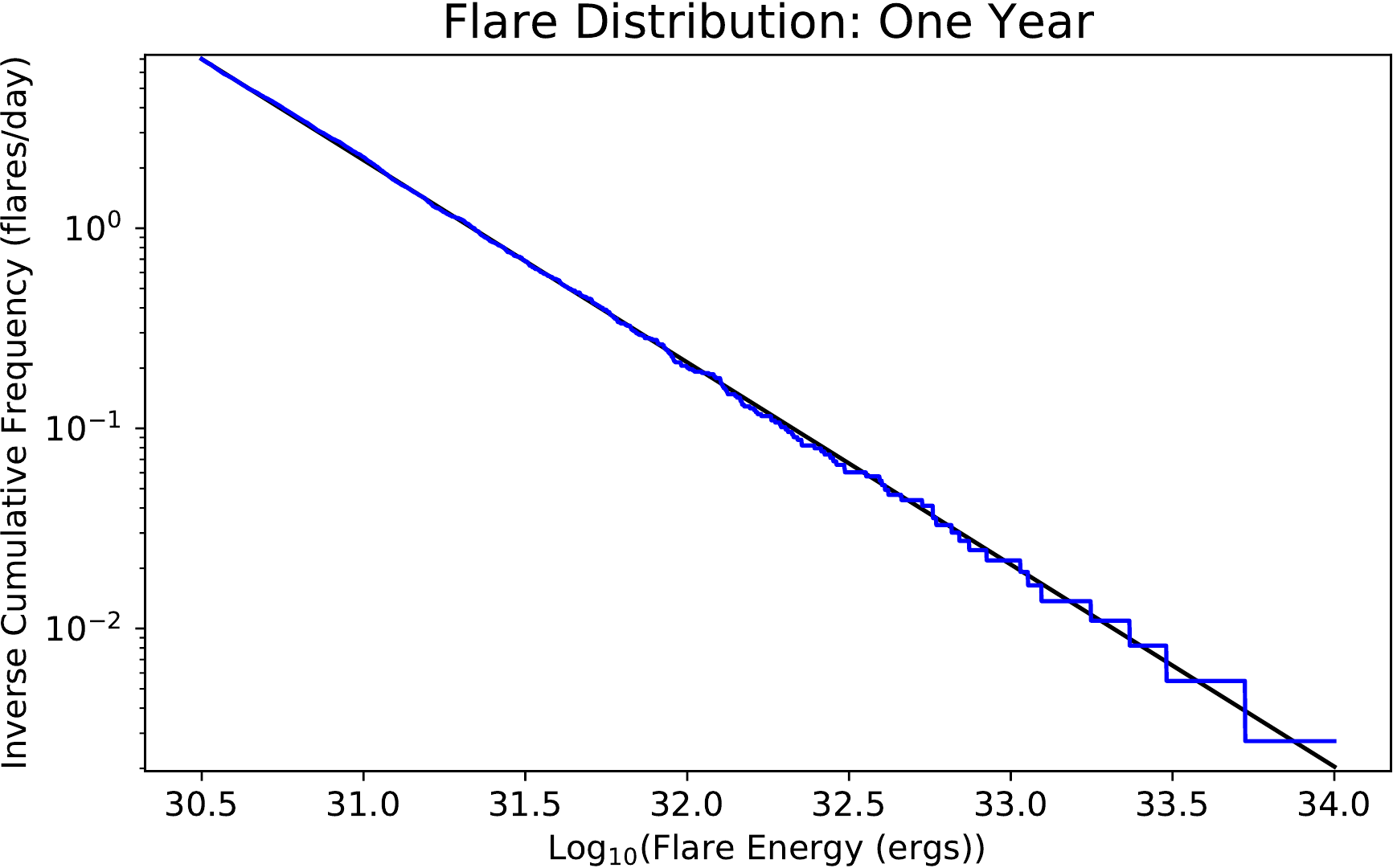} 
    \caption{The inverse--cumulative--frequency (in flares/day) of flare energy from a one year sample (blue line) as compared to the analytic distribution (black line).}
    \label{Fig:FlaringRate}
\end{figure}

In order to capture the impact of repeated flares in our simulations we constructed a sample of flare events that occurred over the course of a year (a `year' refers to an Earth year). \citet{Hawley2014} and \citet{Davenport2014} found that flares on M dwarfs range in duration from minutes to hours, and in magnitude of flare energy (the energy from the increased electromagnetic radiation) from $10^{28-34}$\,ergs (1 erg = $10^{-7}$\,J). We used the occurrence-flare energy distribution from \citet{Tilley2019}, derived from Kepler observations of the M dwarf GJ~1243 \citep{Hawley2014} given as,
\begin{equation}
    \log_{10} \nu = -1.01 \log_{10} E + 31.65, \label{Eqn:InvCumSum}
\end{equation}
where $\nu$ is the inverse-cumulative frequency of the flares (flares/day) and $E$ is the energy of the flare in ergs. The increased temporal resolution (see Table~\ref{Tab:sim_names}) of simulations with flares\footnote{The simulations run approximately 50x slower than those without flares} at this stage meant that simulations were only run for a single year. However, the sample of flares from a period of a single year could have significant inter--annual variability, and differ considerably from the analytic distribution. A sample of flares from a one year period was  generated $\approx10^6$ times and the sample which best matched the distribution was used in this work (i.e. the sample of flares with the minimum $l^2$-norm when comparing the occurrence-flare energy distribution from the sample to the analytic distribution) shown in Figure~\ref{Fig:FlaringRate}. We note that that choice was made to force the one year samples to include a $10^{34}$\,erg flare with an associated CME that occurs 60\,days into the simulation. This was done in order to observe the impacts of a maximum strength flare and CME, the results of which are discussed in Section~\ref{subsec:flaring}. 

We constructed a time--varying stellar irradiance that represents a flare, that is applied whenever a flare occurs. We chose the flare template of \citet{Venot2016} representing the 'great flare on April 12th 1985' of AD Leonis (AD Leo), reported and characterised by \citet{Hawley1991}, and used this to alter the input stellar irradiance. The flare template used by \citet{Venot2016} modifies the irradiance between 100 and 444\,nm only, in future this could be extended to longer wavelengths, however, the increased flux from the flaring decreases rapidly beyond 444\,nm. The flare template was converted so the flare would represent a flare occurring on ProxCen. From the AD Leo flare template we obtained scaling factors of the irradiance for the duration of the flare using
\begin{equation}
    F_{\text{Prox,Flare}}(\lambda,t) = F_{\text{Prox,Qui}}(\lambda,t) \frac{F_{\text{ADLeo,Flare}}(\lambda,t)}{F_{\text{ADLeo,Qui}}(\lambda,t)},
\end{equation}
where $F_{\text{Prox,Qui}}$ and $F_{\text{Prox,Flare}}$ are the stellar irradiances of ProxCen during quiescent conditions and during a given stellar flare respectively, and $F_{\text{ADLeo,Qui}}$ and $F_{\text{ADLeo,Flare}}$ are the stellar irradiances of AD Leo during quiescent conditions and during a given stellar flare, respectively.

\begin{figure}
    \centering
    \includegraphics[width=\columnwidth]{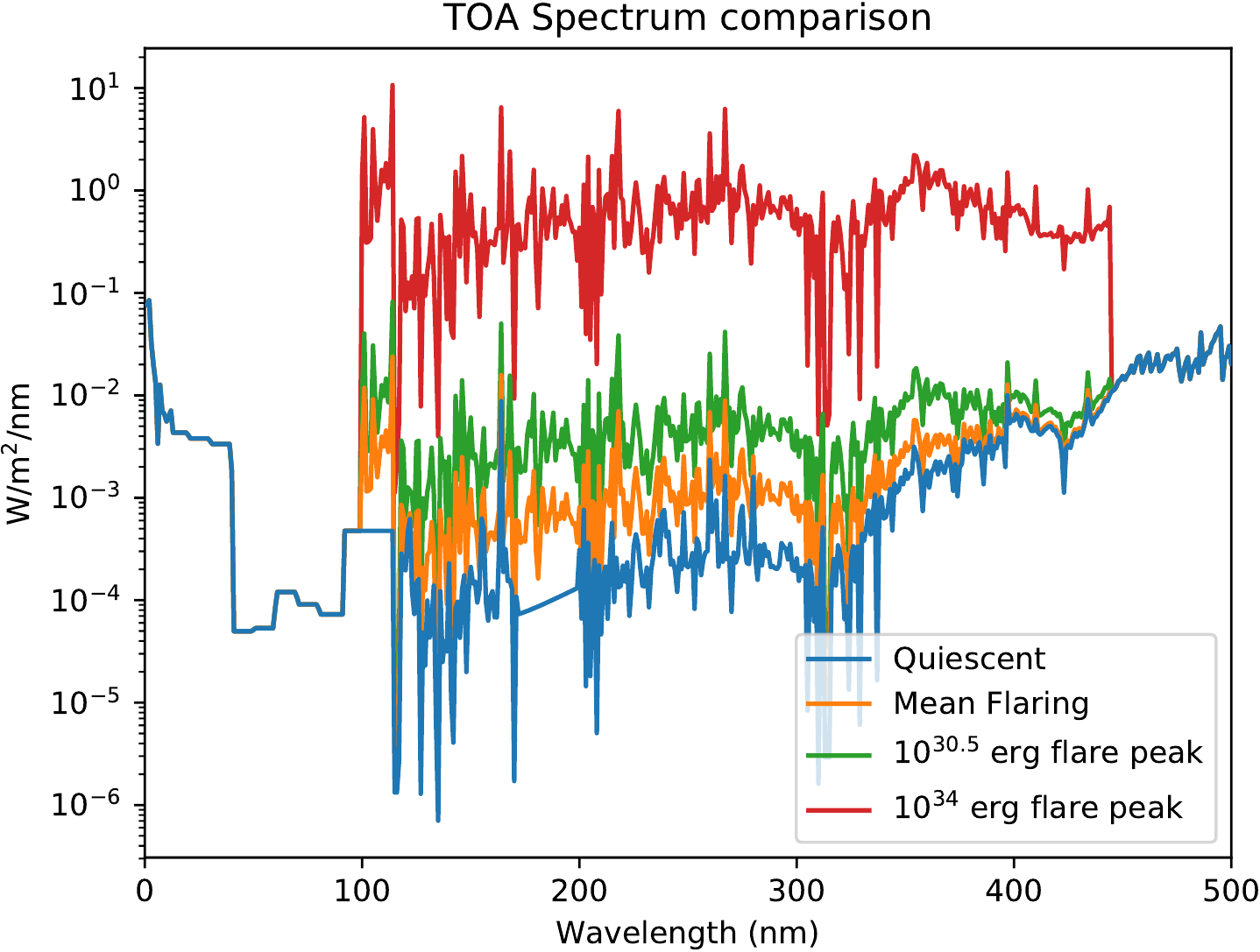}
    \caption{The top--of--atmosphere 'mean flaring' stellar spectrum compared to the quiescent stellar spectrum and the spectrum at the peak of a $10^{30.5}$ and $10^{34}$\,erg flare. The 'mean flaring' spectrum was created by calculating the mean spectrum over the year of simulated flares.}
    \label{Fig:FlaringSpectra}
\end{figure}

The duration of the flares was assumed to follow the power law from \citet{Tilley2019}, derived from the observations of flares reported by \citet{Hawley2014}, namely,
\begin{equation}
    \log_{10} t = 0.395 \log_{10} E - 9.269,
\end{equation}
where $t$ is the duration of the flare in seconds and $E$ is the energy released by the flare in ergs. This was applied by scaling the template spectrum in duration and magnitude. The magnitude of the flare was scaled so that the energy released during the flare was consistent with that indicated by our flare distribution. Figure~\ref{Fig:FlaringSpectra} shows the quiescent spectrum, the spectra at the peak of a $10^{30.5}$ and $10^{34}$\,erg flare, and a `mean flaring' spectrum calculated as the time mean over the flaring period. The CME profile, and impact probability are discussed in Section \ref{subsec:protons}. As Figure~\ref{Fig:FlaringSpectra} shows the quiescent and flaring spectra diverge significantly between 100-444\,nm. The `mean flaring' spectrum is much weaker than the peak $10^{34}$\,erg flare spectrum, as expected. This tells us that the atmosphere's response to a `mean flaring' spectrum will diverge from the peak spectrum significantly, due to the much higher UV radiation driving higher photolysis rates. The impacts of the resolved flares are examined in Section~\ref{subsec:flaring}.

\subsection{Chemistry}
\label{subsec:chemistry}
This section describes the extension of the chemical kinetics framework used in \citet{Drummond2020} and the SOCRATES radiative transfer scheme to include photolysis, a parameterisation of the stellar proton flux (and the reactions that the protons cause in the atmosphere), chemical feedback of water vapour, and dry deposition. Finally, we describe the networks of chemical reactions used in the simulations.

\subsubsection{Photolysis}
\label{subsec:photolysis}
Photolysis rates are calculated using a treatment developed for the UM's radiative transfer scheme SOCRATES as part of efforts to model the effects of space weather, such as geomagnetic storms, solar flares, and CMEs \citep[see][for more details]{Jackson2020}. This implementation includes a treatment of the spherical geometry for the radiation, and the impacts of radiation at UV and X-ray wavelengths for heating and photolysis. We have also explored the impact of spherical geometry on the clouds on GJ~1214b, comparing results both with and without the spherical geometry treatment in Christie et al., (under review). In this study, photolysis includes photodissociation directly caused by radiation, and secondary dissociations caused by photoelectrons (free electrons released by photoionisations), for the wavelength range in this study, namely, 0.5\,nm-10\,$\mu$m. 



The \citet{Drummond2020} chemistry framework did not include photolysis. The SOCRATES photolysis scheme was coupled to the chemistry framework. During a chemical time--step, the photolysis scheme generates photolysis rates which are then used in the chemistry scheme to model changes in atmospheric composition due to photolysis. The modified molecular abundances are then input into the chemical kinetics solver \citep{Drummond2020}, aside from the amount of water vapour which is mostly controlled by the UM's microphysics scheme \citep{Wilson2008}, but is impacted by \ce{H2O} photolysis and SEPs. This is described in Section \ref{subsec:water_vapour}. We do not include any surface emissions and have no surface boundary conditions for the concentration of any chemical species.  The updated atmospheric composition calculated by the chemistry scheme are then passed back to SOCRATES to control atmospheric heating and photolysis. A full list of the photolysis reactions, and species we track, is included in Appendix \ref{appsec:reactions}.

\subsubsection{Stellar Proton Forcing}
\label{subsec:protons}
To parameterise the effects of stellar protons (or stellar energetic particles, SEPs), we used ionisation rates observed in Earth's atmosphere, and scaled them with ProxCen flare strength. We take a similar approach to previous work \citep{Chen2020}, using the solar proton data provided for CMIP6\footnote{obtainable from SOLARIS-HEPPA\\ \url{https://solarisheppa.geomar.de/solarprotonfluxes}}. These data consist of proton fluxes measured from various space--based instruments and provides ionisation rates in the atmosphere due to solar protons. The rates are used to determine the reaction rates of the following reactions\begin{chemequations}
    \begin{alignat}{4}
        \ce{H2O &-> H + OH}, \label{Reac:proton1}\\
        \ce{N2 &-> N(^4S) + N(^4S)}, \label{Reac:proton2}\\
        \ce{N2 &-> N(^2D) + N(^2D)}, \label{Reac:proton3}
    \end{alignat}
\end{chemequations} \\
where \ce{N(^4S)} is ground state atomic nitrogen, and \ce{N(^2D)} is an excited state of atomic nitrogen. Following \citet{Solomon1981a} and \citet{Porter1976}, we assume that 2 \hox molecules (one \ce{H} and one \ce{OH} molecule) and 1.25 nitrogen atoms (0.7 \ce{N(^2D)} and 0.55 \ce{N(^4S)} atoms) are created per ionisation. The full list of reactions included in our model are given in Appendix \ref{appsec:reactions}.

Figure~\ref{Fig:IonisationRates} shows examples for the ionisation rates under quiescent conditions, a minimum strength CME (from a 10$^{30.5}$\,erg flare), and a maximum strength CME (from a 10$^{34}$\,erg flare), used in our model. Two ionisation profiles are used in this work, a quiescent profile used when a CME is not impacting the planet (this can be thought to represent the stellar wind) and a scalable CME profile used when a CME is impacting the planet. The quiescent profile was created from the mean of ionisation data from 2009 (around the time of solar minimum). The CME profile was created using data from the peak of the 2003 solar storm. To determine how the rates change with flare strength, we use the relation of flare strength to proton flux from \citet{Belov2005} which relates proton flux to the peak X--ray intensity during the flare. As described in \citet{Tilley2019} the proton flux (in proton flux units, or pfu) is given by
\begin{equation}
    I_p(>10\text{MeV}) = k_0 \left(\frac{\phi^{1.08}}{a^2}\right)^{1.14},
\end{equation}
where $\phi$ is the relative flux increase in the Johnson U band, $a$ is the semi-major axis in astronomical units. We represent the transient nature of CMEs by only applying the scalable CME profile during the flare (if a CME impacts the planet), otherwise the planet only receives the quiescent profile. This is applied as a constant for the duration of the flare, with no correction for a delay between the onset of the flare and the onset of the CME. Solar CMEs range in velocities from $\sim30-2600$\,km s$^{-1}$, with an average velocity of 428\,km s$^{-1}$ \citep{Yashiro2004}. For our simulations assuming the planet is orbiting at a distance of 0.0485\,AU, that would mean an average delay of $\sim4.7$\,hours, ranging from $\sim0.77$-$67.2$\,hours. The response of the atmosphere when the CMEs are delayed is a topic for future research.

\begin{figure}
    \centering
    \includegraphics[width=\columnwidth]{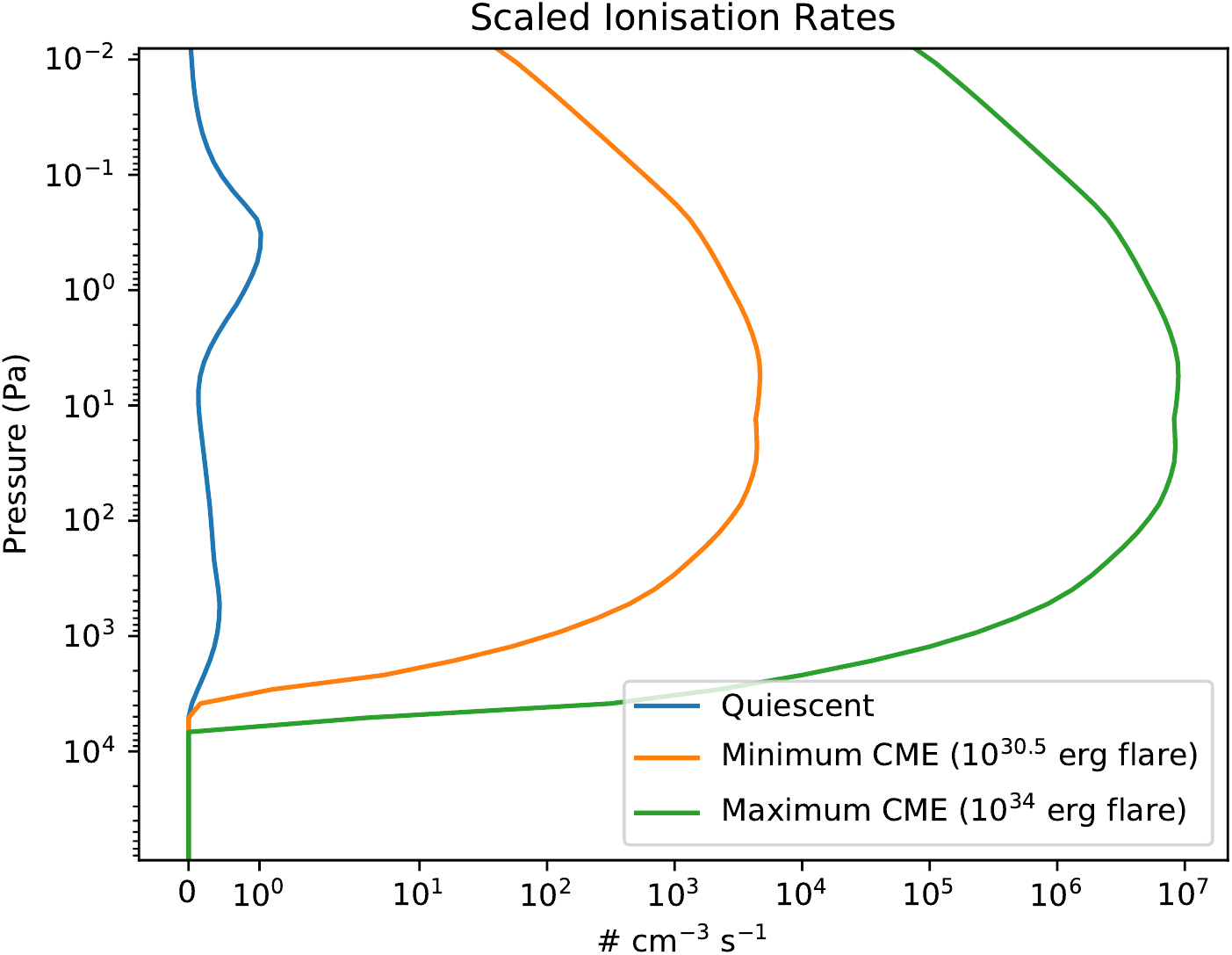}
    \caption{The ionisation rates during quiescent conditions (blue), the ionisation rates during a 10$^{30.5}$ erg flare (orange), and a 10$^{34}$ erg flare (green).}
    \label{Fig:IonisationRates}
\end{figure}

In order to determine whether protons from a given CME impact the day--side of our planet, we have calculated the impact probability, $P$, using the formula given by \citet{Khodachenko2007} which accounts for the geometry of the CME, the size of the CME, and where on the star the CME originated. The formula is
\begin{equation}
    P=\frac{(\Delta_\text{CME} + \delta_\text{pl}) \sin[(\Delta_\text{CME} + \delta_\text{pl})/2]}{2\pi \sin(\Theta)},
\end{equation}
where $\Delta_\text{CME}$ is the angular width of the CME, $\delta_\text{pl}$ is the angular size of the planet, and $\Theta$ is the interval of stellar latitudes (from -$\Theta$, to $\Theta$) where CMEs can originate from. Assuming values of $\Delta_\text{CME}=47^\circ$ \citep{Gopalswamy2004}, $\Theta=60^\circ$ \citep{Gopalswamy2008a}, and $\delta_\text{pl} \approx 10^{-3}$ (the angular size of a planet with a 7160\,km radius seen from a distance of 0.0485\,AU), we find a CME impact probability of 0.06. This probability is applied to every flare to create the CMEs used in this simulation. This probability is quite low, but does result in the planet being hit by a CME every few days on average.

In this work we assume that every flare has an associated CME. This is similar to \citet{Tilley2019} and \citet{Chen2020}. However, recent work by \citet{Muheki2020,Alvarado-Gomez2021} indicates that the relationship between solar flares and CMEs is not appropriate for M dwarf stars, where there is less CME activity. Therefore, our results will effectively over--emphasise the effects of CMEs.

\subsubsection{Dry deposition}
\label{subsec:deposition}
In addition to the direct impacts of increased UV radiation and protons, species involved in ozone chemistry can also be removed from the atmosphere via `dry' and `wet' deposition. In this work, we have not included wet deposition (which will result in a high bias in the concentrations of species such as nitric acid (\ce{HNO3}), dinitrogen pentoxide (\ce{N2O5}), and peroxynitric acid (\ce{HO2NO2})), but do include a treatment of dry deposition. Dry deposition is a process whereby particles and trace gases settle onto the planetary surface, either by turbulence, or through gravitational settling (although this is only relevant for large particles) . This acts to remove the affected species from the atmosphere. In this work dry deposition was implemented in the simple form described by \citet{Giannakopoulos1999} for the following seven species: \ce{O3}, \ce{NO2}, \ce{NO3}, \ce{N2O5}, \ce{HO2NO2}, \ce{HNO3}, and \ce{H2O2}. We chose to parameterise dry deposition through a single deposition velocity $V_{dep}$. $V_{dep} \equiv F_c/n_i$, is the ratio of the flux density of particles ($F_c$) onto the surface to the number density of particles in the air above the surface ($n_i$). $V_{dep}$ varies for different terrains \citep{Giannakopoulos1999}, but as we are simulating an aquaplanet we adopt values of $V_{dep}$ representing dry deposition onto an ocean which is used for the entire planetary surface. Table~\ref{Tab:Deposition} lists the deposition velocities used for each species. From the definition of $V_{dep}$ we can construct a first-order differential equation modelling the change in number density of affected species
\begin{equation}
    \pd{n_{i}}{t} = -\frac{V_{dep} A}{V} n_i,
\end{equation}
where $V_{dep}$ is the deposition velocity of species $i$, and $A$ and $V$ are the surface area and volume of the bottom level grid cells used in the simulations. More advanced dry deposition schemes determine the deposition velocity using a resistance-based approach \citep{Wesley1989} accounting for several factors such as the winds near the surface and the effect of the type of surface (ocean, forests, urban environments, etc.) to accept the molecule. Incorporating this into our model is the aim of future work.

\begin{table}
\caption{The deposition velocities for the species affected by dry deposition in this work.}
\label{Tab:Deposition}
\begin{tabular}{lll}
Species                     & $V_{dep}$ at 1\,m from surface (cm s$^{-1}$) \\ \hline  \hline
\ce{O3}                     & 0.05                             \\
\ce{NO2}                    & 0.02                             \\
\ce{NO3}                    & 0.02                             \\
\ce{N2O5}                   & 1.00                             \\
\ce{HO2NO2}                 & 1.00                             \\
\ce{HNO3}                   & 1.00                             \\
\ce{H2O2}                   & 1.00                                     
\end{tabular}
\end{table}

\subsubsection{Water vapour feedback}
\label{subsec:water_vapour}
Water vapour plays a key role in \hox chemistry and is a species destroyed by SEPs (see Reaction \ref{Reac:proton1}) which, in turn, has a significant impact on the ozone abundance. Water through direct opacity and the formation of clouds, also plays a key role in determining the overall climate of the planet. Therefore, in order to capture the water abundance more accurately, we have also incorporated water vapour feedback into the chemistry model. While the main production of water vapour is through evaporation from the surface \citep[see][for details]{Boutle2017} and is still controlled by the climate model, we also include destruction and production of water vapour in the chemical network. Inclusion of this additional element of consistency in the water abundance, however, was not found to significantly change the amount of water vapour present in the atmosphere except at very high altitude (as a result of \ce{H2O} photolysis).

\subsubsection{Chemical networks}
\label{subsec:networks}
We constructed chemical networks (a list of chemical reactions) that describe the Chapman cycle, the \hox catalytic cycle, and the \nox catalytic cycle. The Chapman cycle consists of a series of reactions between molecular oxygen, atomic oxygen, and ozone, with the most important reactions summarised as\begin{chemequations}
    \begin{alignat}{3}
        \ce{O2 &+ h{\nu} & & -> O(^3P) + O(^3P)}, \\
        \ce{O(^3P) + O2 &+ M & &-> O3 + M}, \\
        \ce{O3 &+ h{\nu} & & -> O2 + O(^3P)}, \\
        \ce{O2 &+ h{\nu} & & -> O(^3P) + O(^1D)}, \label{Reac:HOx1}\\
        \ce{O3 &+ h{\nu} & & -> O2 + O(^1D)}, \\
        \ce{O3 &+ O(^3P) & & -> O2 + O2}.
    \end{alignat}
\end{chemequations}
The chemical network describing the Chapman cycle has 6 chemical species, 6 bimolecular reactions, 1 termolecular reaction, and 4 photolysis reactions. 

The \hox catalytic cycle is started by the destruction of water vapour via several mechanisms, an interaction with excited atomic oxygen produced by oxygen/ozone photolysis, and through the photolysis of water vapour. This produces \ce{OH}, which depletes ozone by the following reactions\begin{chemequations}
    \begin{alignat}{3}
        \ce{H2O &+ O(^1D) & & -> OH + OH},  \\
        \ce{H2O &+ h{\nu} & & -> H + OH}, \\
        \ce{OH &+ O3 & & -> HO2 + O2},\\    
        \ce{HO2 &+ O3 & & -> OH + O2 + O2},  \label{Reac:HOx2}\\
        \ce{OH &+ HO2 & & -> H2O + O2}.
    \end{alignat}
\end{chemequations}

The network describing the Chapman cycle and the \hox cycle has every reaction and chemical species from the previous network, as well as 5 additional chemical species, 16 bimolecular reactions, 3 termolecular reactions, and 7 photolysis reactions. In the presence of SEPs, additional \hox is generated through \ce{H2O -> H + OH}. 

The \nox cycle behaves similarly. The \nox cycle depletes ozone via the following reactions\begin{chemequations}
    \begin{alignat}{3}
        \ce{NO + &O3 &-> &NO2 + O2}, \label{Reac:NOx1} \\
        \ce{NO2 + &h{\nu} &-> &NO + O(^3P)}, \label{Reac:NOx2}  \\
        \ce{O(^3P) + &NO2 &-> &NO + O2}. \label{Reac:NOx3} 
    \end{alignat}
\end{chemequations}\\
Without any SEPs, \ce{NO} is generated by\begin{chemequations}
    \begin{alignat}{3}
        \ce{N2 + O(^1D) + &M &-> &N2O + M}, \\
        \ce{N2O + &O(^1D) &-> &NO + NO}.
    \end{alignat}
\end{chemequations}\\
With SEPs, \ce{NO} is modified by the generation of atomic nitrogen
\begin{chemequations}
    \begin{alignat}{3}
        \ce{N2 &-> N(^4S) + N(^4S)},\\
        \ce{N2 &-> N(^2D) + N(^2D)},\\
        \ce{O2 + N(^4S) &-> NO + O(^3P)},\\
        \ce{O2 + N(^2D) &-> NO + O(^3P)},\\
        \ce{NO + N(^4S) &-> N2 + O(^3P)}.
    \end{alignat}
\end{chemequations}\\
The network describing the Chapman cycle, the \hox cycle, and the \nox cycle contains every previously described reaction and chemical species and an additional 10 chemical species, 24 bimolecular reactions, 7 termolecular reactions, and 10 photolysis reactions. All networks also include the dry deposition of every possible species that is included in their network. Two additional networks were created that neglect SEPs, in order to test how important the quiescent stellar wind might be for the generation of \nox molecules. For more details listing all chemical reactions, all species involved in the chemistry, and the coefficients used for each reaction, see Appendix~\ref{appsubsec:bimole}--\ref{appsubsec:sep_reactions}.

\section{Results}
\label{sec:results}
In this work we present results from nine simulations, which we separate into a `Quiescent' phase (five simulations, discussed in Section \ref{subsec:quiet}) and a `Flaring' phase (four simulations, discussed in Section \ref{subsec:flaring}). The impact of quiescent SEPs during the first 12000\,days of the simulations is discussed in Section~\ref{subsec:quiet}, Table \ref{Tab:sim_names} lists all of the simulations we have performed for this work, the stellar irradiation and whether it includes flares, whether the SEP ionisation is quiescent or includes CMEs and the time-steps for the dynamics, radiation and chemistry for each simulation. To verify that the simulations are stable for long periods of time, an initial simulation was run for 3000 Earth days without chemistry, and the fixed `Earth--like' atmospheric composition as prescribed in Table 2 of \citet{Boutle2017}. The end point of this simulation, which was in a climatic steady--state (near constant mean surface temperatures and top--of--atmosphere radiative flux balance) was then used as the start point for the quiescent phase simulations with chemistry. Likewise, the end state of the quiescent simulation containing the full chemical network and quiescent SEPs was used as the start point for the flaring simulations.
\begin{table*}
\caption{All simulations performed for this work, with short names, description of the components included and the timesteps used. See text for details of the input stellar spectra. The full chemical reactions are detailed in Appendix \ref{appsec:reactions} and the profiles for the spectra and SEPs described in Sections \ref{subsec:spectra} and \ref{subsec:protons}, respectively.}
\label{Tab:sim_names}
\begin{tabular}{llllllll}
\multirow{2}{*}{Phase}&\multirow{2}{*}{Name}&\multirow{2}{*}{Spectrum}&\multirow{2}{*}{Chemistry}&\multirow{2}{*}{SEPs (affected species)}&\multicolumn{3}{c}{Timesteps (minutes)}\\
&&&&&Dynamic&Radiation&Chemistry\\
\hline  \hline
\multirow{5}{*}{Quiescent}&Quiet\_Ch&Quiescent&Chapman cycle&--&\multirow{5}{*}{10}&\multirow{5}{*}{60}&\multirow{5}{*}{60}\\&Quiet\_Ch\_HOx&Quiescent&Chapman cycle \& \hox&--&\\
&Quiet\_Ch\_HOx\_SEP&Quiescent&Chapman cycle, \hox&Quiescent (\ce{H2O})&\\
&Quiet\_Ch\_HOx\_NOx&Quiescent&Chapman cycle, \hox \& \nox&--&\\
&Quiet\_Full&Quiescent&Chapman cycle, \hox \& \nox&Quiescent (\ce{H2O, N2})&\\
\hline
\multirow{3}{*}{Flaring}& Control&Quiescent&Chapman cycle, \hox \& \nox & Quiescent (\ce{H2O, N2}) &\multirow{4}{*}{2}&\multirow{4}{*}{2}&\multirow{4}{*}{2}\\
& Flare\_UV&Flaring&Chapman cycle, \hox \& \nox& Quiescent (\ce{H2O, N2}) &\\
& Flare\_Full & Flaring&Chapman cycle, \hox \& \nox & Flaring (\ce{H2O, N2})&\\
& Mean\_Flare & Mean flaring & Chapman cycle, \hox \& \nox & Quiescent (\ce{H2O, N2}) &\\
\hline
\end{tabular}
\end{table*}

For the quiescent phase, the planet was simulated for 12000\,days under quiescent conditions (green line Figure~\ref{Fig:Spectra}) but adopting five different configurations, all employing the timesteps given in the last three columns of Table \ref{Tab:sim_names}. These configurations covered a range of chemical networks, starting with only the Chapman cycle included (Quiet\_Ch), then adding the \hox chemistry both without (Quiet\_Ch\_HOx, differing from \citet{Yates2020} by the inclusion of \ce{H}, \ce{H2O2}, and \ce{H2O} photolysis), and with (Quiet\_Ch\_HOx\_SEP) quiescent SEPs (as detailed in Section \ref{subsec:protons}), and finally further adding \nox chemistry (i.e. the full chemical network, see Appendix \ref{appsec:reactions} for details) again both without (Quiet\_Ch\_HOx\_NOx)  and with (Quiet\_Full) a quiescent SEP profile. Unlike \citet{Drummond2020} we do not use Gibbs minimisation to determine the initial atmospheric composition, but set the initial values manually. The species included in each quiescent simulation and their initial mass fractions are listed in Table~\ref{Tab:InitialMass}.

\begin{table*}
\caption{The species included in each of the quiescent simulations, and the initial mass fractions of each species. As the water mass fraction was controlled by the UM, its initial value was not constant but was a range of values. Due to this, and the differing number of species in the network, the initial mass fraction of \ce{N2} differs slightly between networks.}
\label{Tab:InitialMass}
\begin{tabular}{lllllll}
Species & Initial mass fraction (kg kg$^{-1}$) & Quiet\_Ch & Quiet\_Ch\_HOx & Quiet\_Ch\_HOx\_SEP & Quiet\_Ch\_HOx\_NOx & Quiet\_Full \\ \hline \hline
\ce{CO2}    & 5.941$\times10^{-4}$ & X & X & X & X & X    \\
\ce{O2}     & 0.2314               & X & X & X & X & X    \\
\ce{O(^3P)} & 0                    & X & X & X & X & X    \\
\ce{O(^1D)} & 0                    & X & X & X & X & X    \\
\ce{O3}     & $10^{-9}$            & X & X & X & X & X    \\
\ce{H2O}    & $10^{-6}-10^{-2}$    & X & X & X & X & X    \\
\ce{N2}     & $\sim0.76$           & X  & X  & X  & X & X \\ \hline
\ce{OH}     & $10^{-12}$           & -- & X & X & X & X   \\
\ce{HO2}    & $10^{-12}$           & -- & X & X & X & X   \\
\ce{H}      & $10^{-12}$           & -- & X & X & X & X   \\
\ce{H2}     & 0                    & -- & X & X & X & X   \\
\ce{H2O2}   & $10^{-12}$           & -- & X & X & X & X   \\ \hline
\ce{N(^4S)} & $10^{-12}$           & -- & -- & -- & X & X \\
\ce{N(^2D)} & $10^{-12}$           & -- & -- & -- & X & X \\
\ce{NO}     & $10^{-12}$           & -- & -- & -- & X & X \\
\ce{NO2}    & $10^{-12}$           & -- & -- & -- & X & X \\
\ce{NO3}    & $10^{-12}$           & -- & -- & -- & X & X \\
\ce{HNO3}   & $10^{-12}$           & -- & -- & -- & X & X \\
\ce{N2O}    & $10^{-12}$           & -- & -- & -- & X & X \\
\ce{N2O5}   & $10^{-12}$           & -- & -- & -- & X & X \\
\ce{HONO}   & $10^{-12}$           & -- & -- & -- & X & X \\
\ce{HO2NO2} & $10^{-12}$           & -- & -- & -- & X & X
\end{tabular}
\end{table*}

The end point of the full quiescent version, i.e. our Quiet\_Full simulation including 12000\,days with the full chemistry and quiescent SEP profile included, was then used as the start point for four further simulations. All four of these simulations included the full chemical network, and were run at much finer temporal resolution (the time-step is given in the last three columns of Table \ref{Tab:sim_names}) for one year (due to the very high computational cost associated with such simulations). These four simulations comprised of a Control which continued under quiescent irradiation and SEPs, then two further simulations irradiated with a flaring spectrum (constructed as discussed in Section~\ref{subsubsec:flare_profile}), one omitting SEPs (Flare\_UV) and another including SEPs (Flare\_Full). Finally, we performed a Mean\_Flare simulation, where the stellar spectrum remains constant at the time--mean of the stellar spectrum over the flaring period used for the other flaring cases (described in Section~\ref{subsubsec:flare_profile}, and Figure~\ref{Fig:FlaringSpectra}), with the SEPs held at the quiescent rates. This was done so we could test the importance of the time-dependent spectra, as using a mean flare spectrum would present a significant increase in computation speed and allow us to examine the response to flares over a longer period of time.

In Section~\ref{subsec:quiet} we first discuss the results of our simulations from the quiescent phase, particularly in comparison to \citet{Yates2020} who also used the UM, before moving onto the flaring simulations in Section~\ref{subsec:flaring}. 

\subsection{Quiescent} 
\label{subsec:quiet}
\begin{figure*}
    \centering
    \includegraphics[width=\textwidth]{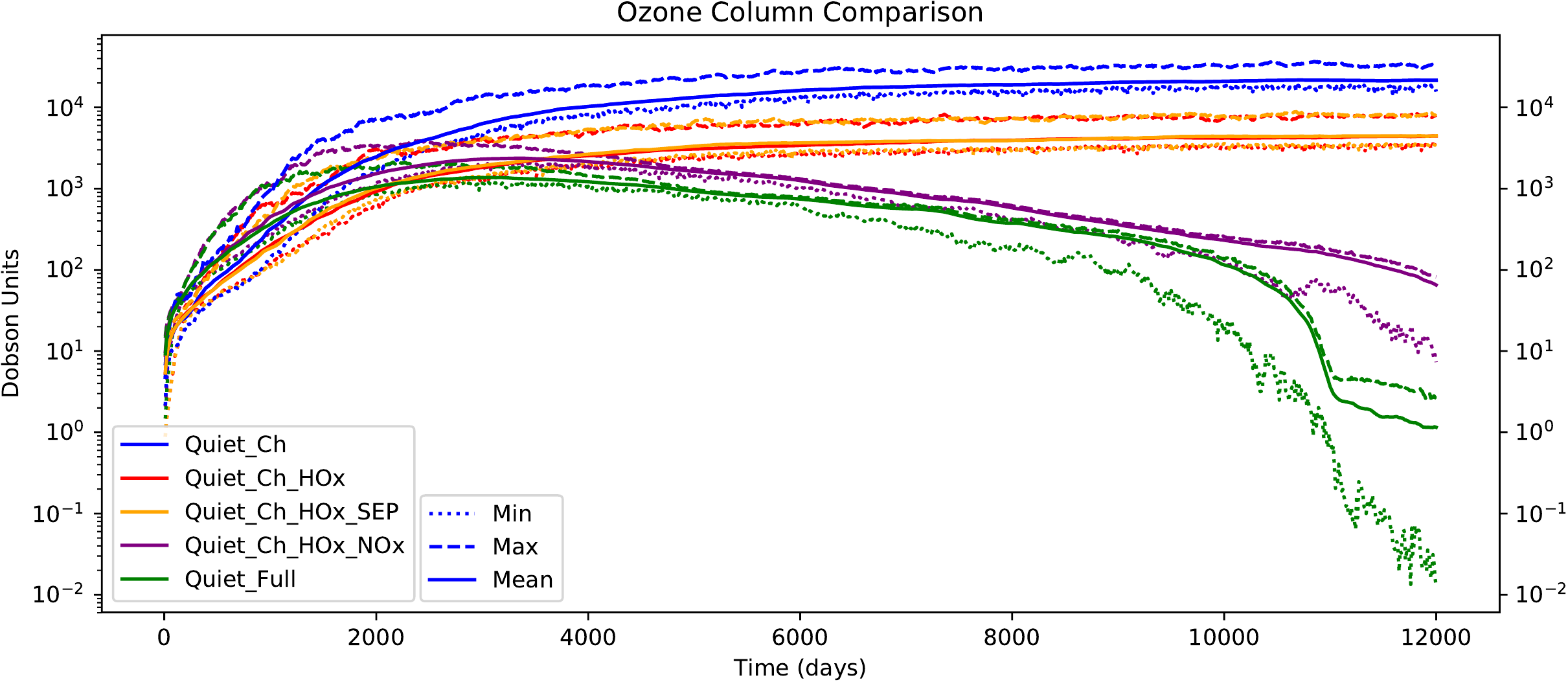}
    \caption{The globally averaged total ozone column measured in Dobson Units for the five chemical networks. Refer to Table~\ref{Tab:sim_names} for the details of the different networks.}
    \label{Fig:OzoneColumnCompLog}
\end{figure*}

\begin{figure}
    \centering
    \includegraphics[width=\columnwidth]{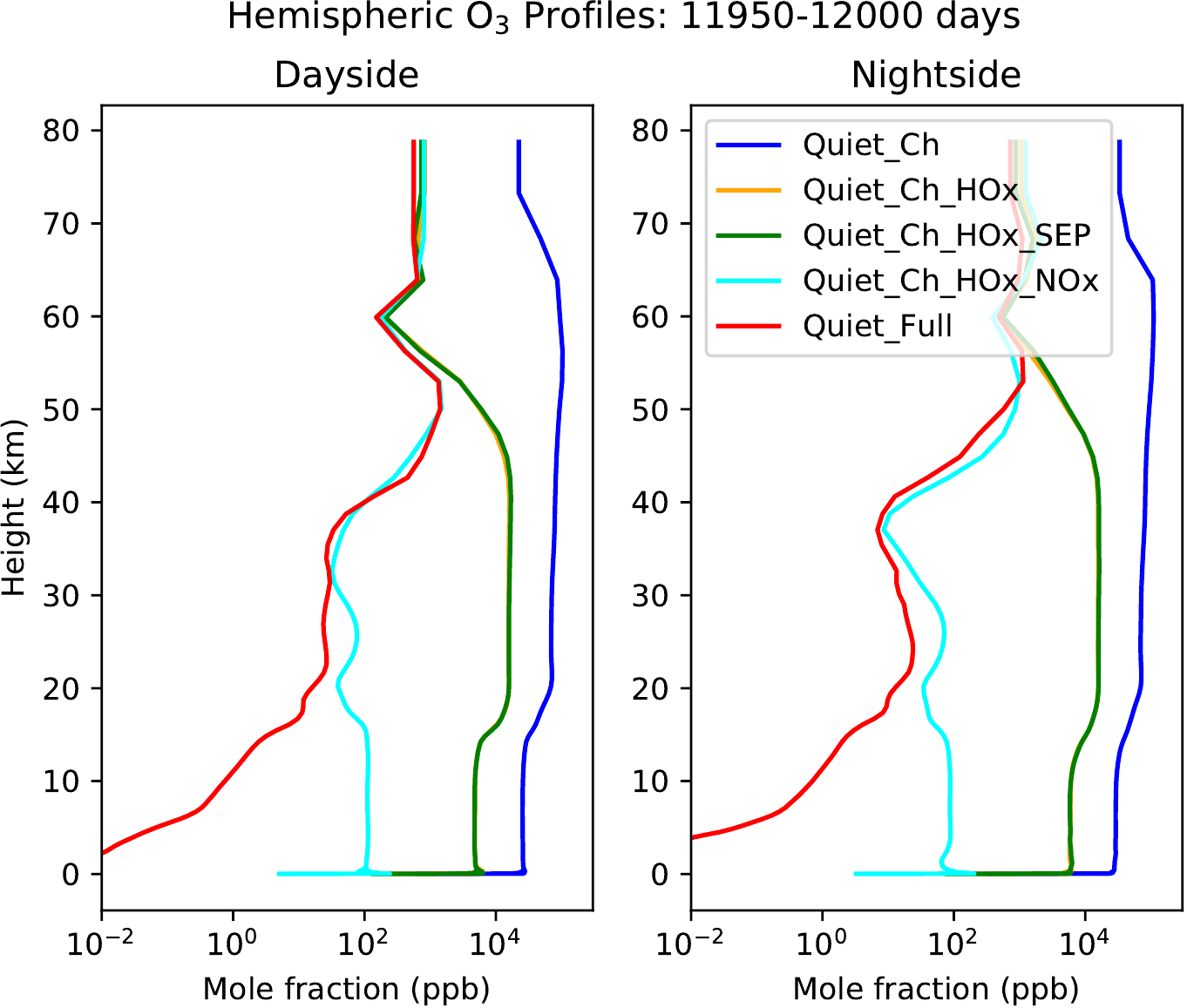}
    \caption{The spatially averaged vertical profile of the ozone mole fractions from the planets day--side and night--side for the five chemical networks under quiescent conditions. The profiles for Quiet\_Ch\_HOx and Quiet\_Ch\_HOx\_SEP almost completely overlap.  Refer to Table~\ref{Tab:sim_names} for the details of the different networks.}
    \label{Fig:VerticalOzoneProfileQuiescent}
\end{figure}

The Quiet\_Ch simulation allows us to isolate the ozone production through the Chapman cycle, and effectively test the model performance. Figure~\ref{Fig:OzoneColumnCompLog} shows the globally averaged ozone column in Dobson Units (DU, 1 DU = 2.69$\times$10$^{20}$\,molecules/m$^2$) for the five different networks in quiescent conditions described in Table~\ref{Tab:sim_names}, as well as illustrating the range of values of the ozone column for each network. We find that the networks behaved as expected. The simplest network consisting only of the Chapman cycle has the largest ozone column ($\approx 22000$\,DU), and the introduction of \hox and \nox chemistry heavily diminishes the ozone column, particularly in the latter case. The introduction of \hox and \nox chemistry heavily depletes tropospheric and stratospheric ozone (through the reactions listed in Section~\ref{subsec:networks}), as Figure~\ref{Fig:VerticalOzoneProfileQuiescent} shows. The stratosphere is more heavily depleted than the troposphere, but due to the increased density of the air in the troposphere, the depletion of the troposphere contributes much more to the changes in the total ozone column. We also find that the inclusion of the quiescent SEP profile does not change the ozone column between the Quiet\_Ch\_HOx and  Quiet\_Ch\_HOx\_SEP simulations appreciatively. We find that SEPs have an impact on the Quiet\_Ch\_HOx\_NOx and Quiet\_Full simulations, the introduction of SEPs reduces the ozone column from $\approx 100$\,DU to $\approx 1$\,DU.

Figure~\ref{Fig:VerticalOzoneProfileQuiescent} shows the spatially averaged day--side and night--side vertical profiles of the ozone mole fraction for the five quiescent simulations. We see that Quiet\_Ch has a substantial amount of tropospheric ozone, $\sim25$\,parts--per--million (ppm) on the day--side and  $\sim 30$\,ppm on the night--side, and a large amount of ozone in the stratosphere, ranging between $\sim 30-110$\,ppm. The introduction of \hox (Quiet\_Ch\_HOx and Quiet\_Ch\_HOx\_SEP) has depleted the tropospheric ozone by $\sim21$\,ppm. Stratospheric ozone has been heavily depleted, now consisting of a layer of $\sim17$\,ppm between $14-50$\,km. Above 55\,km ozone is almost completely depleted, which is attributed to the introduction of \hox and \ce{O3} photolysis being much stronger at high altitudes. The SEPs do not have an appreciable impact on the amount of ozone. The additional \hox generated by Reaction~\ref{Reac:proton1} is quite small compared to the sources of \hox (\ce{H2O photolysis}, \ce{H2O +O(^1D)}), and does not have a significant impact on the depletion of ozone. In contrast, the SEPs have a clearly noticeable impact when we include \nox chemistry (Quiet\_Ch\_HOx\_NOx and Quiet\_Full). The introduction of \nox chemistry has almost completely depleted ozone in the troposphere, Quiet\_Ch\_HOx\_NOx has been reduced to $\sim100$\,parts--per--billion (ppb), and Quiet\_Full has been almost completely depleted, declining from 10\,ppb at 15\,km to less than 10\,parts--per--trillion near the surface. In the stratosphere we see a similarly large depletion, with only a thin layer of $\sim1$\,ppm present between 45-55\,km. The \hox and \nox created by the SEPs (Reactions~\ref{Reac:proton1}-\ref{Reac:proton3}) do not have a large impact, but it is more apparent than the difference between Quiet\_Ch\_HOx and Quiet\_Ch\_HOx\_SEP. We see that Quiet\_Ch\_HOx\_NOx and Quiet\_Full diverge drastically below 40\,km. This divergence is linked to the change in chemistry between the two simulations, likely initiated by SEPs in the mid-stratosphere. Our lack of wet deposition (and heterogeneous chemistry) has resulted in a high bias in \nox reservoirs (\ce{HNO3}, \ce{N2O5}, and \ce{HO2NO2}) in the lower atmosphere. This has heavily increased the amount of reactive nitrogen which was made available in larger amounts by the inclusion of SEPs. In future work, we plan to examine these differences in more detail.

\begin{figure*}
    \centering
    \includegraphics[width=\textwidth]{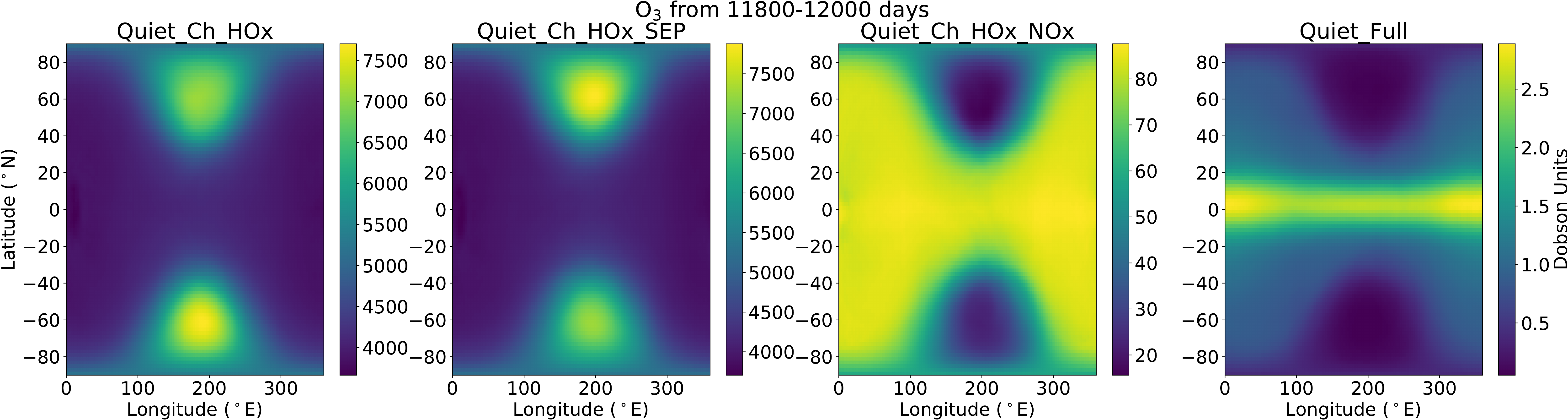}
    \caption{The spatial ozone distribution temporally averaged over the period of 11800-12000\,days for the Quiet\_Ch\_HOx (left), Quiet\_Ch\_HOx\_SEP (centre left), Quiet\_Ch\_HOx\_NOx (centre right) and Quiet\_Full (right) simulations from the quiescent phase in Dobson Units. See Table~\ref{Tab:sim_names} for explanation of the simulation names. The ozone distribution is significantly reduced by the addition of \nox chemistry, and is further depleted by the inclusion of SEPs.}
    \label{Fig:HOxNOx}
\end{figure*}

Figure~\ref{Fig:HOxNOx} shows the distribution of ozone (in DU) for our Quiet\_Ch\_HOx (left), Quiet\_Ch\_HOx\_SEP (centre left), Quiet\_Ch\_HOx\_NOx (centre right) and Quiet\_Full (right) simulations. Similar to previous works our simulations show that the night--side cold--traps can store significant volumes of ozone. Comparing the leftmost panel of Figure~\ref{Fig:HOxNOx} with Figure 4 (top left panel) of \citet{Yates2020} reveals the differences caused by updating the stellar spectrum and, to a lesser extent, the inclusion of shorter wavelengths in the treatment of photolysis, as these are the main differences between our treatment for the Quiet\_Ch\_HOx simulation and that of \citet{Yates2020}. The update to the MUSCLES-Ribas stellar spectrum leads to much higher levels of ozone, however the change is mainly due to the increased UV flux as this result was not found when using the BT-Settl spectrum (not shown). This illustrates the importance of including low (<200\,nm) wavelength fluxes into chemistry models. This effect occurs regardless of whether quiescent SEP forcing is included (left two panels). When including \nox chemistry in our simulations (Quiet\_Ch\_HOx\_NOx \& Quiet\_Full), the ozone distribution changes drastically (the right panels of Figure~\ref{Fig:HOxNOx}), reducing significantly across the entire planet. Ozone is further depleted by the inclusion of SEPs, reducing the global average ozone column from $\sim60$\,DU to $\sim1$\,DU. Ozone is further depleted in the polar regions, and the night--side gyres (cold traps at high latitudes).

\subsection{Flaring}
\label{subsec:flaring}
The main focus of this work is to explore the impact of a flaring M dwarf star on the climate of an `Earth-like' terrestrial, tidally locked, exoplanet. Therefore, in this section we focus on our three simulations including flares, namely Flare\_UV (full chemistry but omitting CMEs), Flare\_Full (full chemistry and CMEs) and Mean\_Flare (full chemistry, a constant mean flaring spectrum and quiescent levels of SEPs), using our Control (full chemistry but quiescent spectrum and SEP profile) as a reference.

\begin{figure*}
    \centering
    \includegraphics[width=\textwidth]{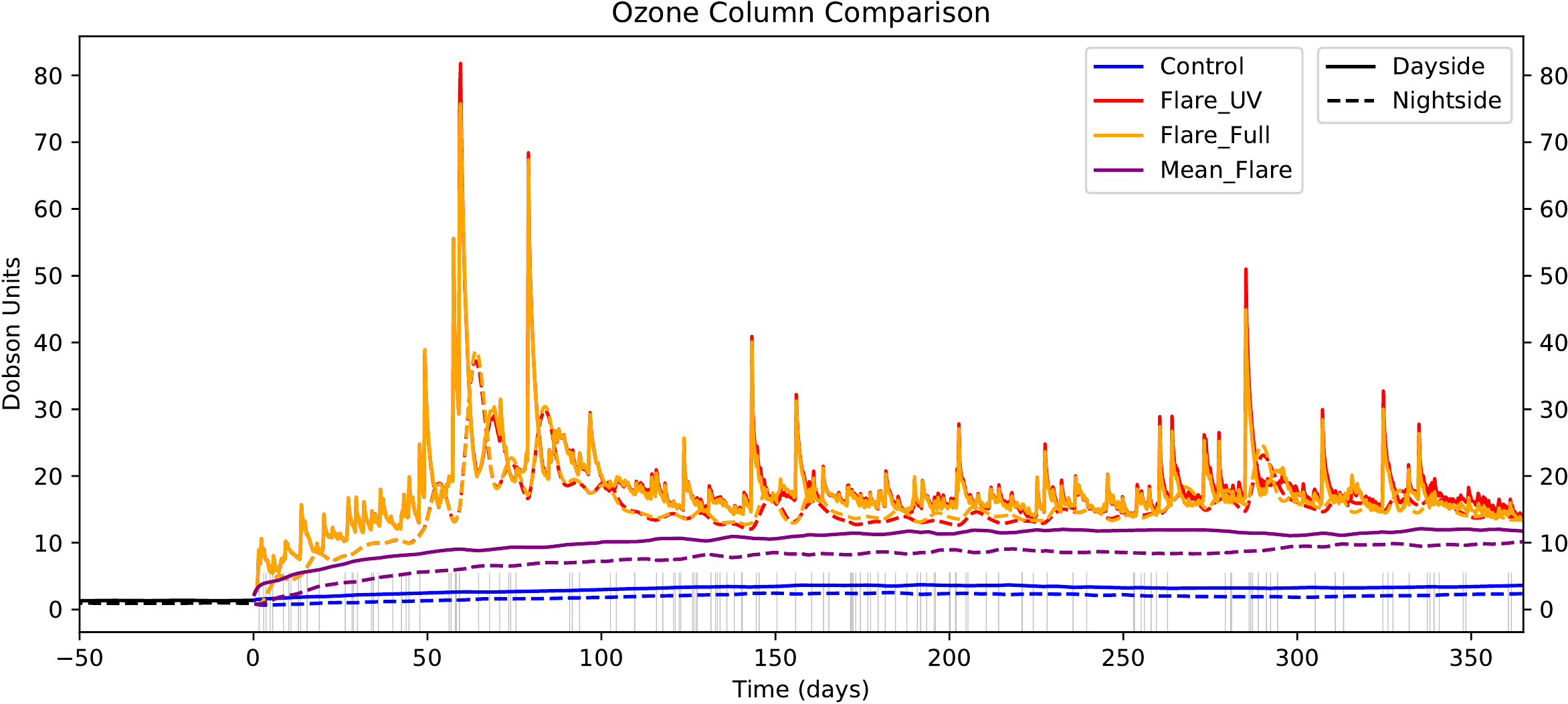}
    \caption{The hemispherically averaged mean ozone column of the day--side and night--side from the flaring simulations used in this work as described in Table~\ref{Tab:sim_names}. CME onset times have been marked in gray.}
    \label{Fig:OzoneDUFlares}
\end{figure*}

Figure~\ref{Fig:OzoneDUFlares} shows the day--side (solid) and night--side (dashed) mean ozone columns during the flaring section of the simulation for the four cases, as well as CME onset times. We find that the impact of the stellar flare irradiation we have constructed is to increase the global averaged total column of ozone in the atmosphere from $\approx 1$\,DU (at the end of Quiet\_Full) to $\approx 15-20$\,DU (the range of values over the last 50 days of the simulations with flares), as the day--side and night--side columns have similar values. During a flare the UV increases substantially and increases \ce{O2} photolysis generating additional atomic oxygen. The increase in atomic oxygen drives the growth of ozone via \ce{O2 + O(^3P) + M -> O3 + M}. The destruction of ozone by photolysis or through additional \hox and \nox increases during a flare as well, but does not increase enough to offset the significant growth in the production of ozone. The net effect of the flare causes the amount of ozone on the day--side to increase rapidly (the largest flares capable of creating an ozone column of >75\,DU), and slowly decrease once the flare has ended. This is due to ozone on the day--side being destroyed and being transported onto the night--side. The enhancement of night--side ozone is due to the advection of ozone from the day--side. The difference in the peak day--side ozone column and peak night--side ozone column (with the flare and CME on day 60 being a good example, where the day--side ozone column peaked at 75-80\,DU, while the night--side column peaked at 40\,DU) demonstrates that the majority of ozone created during a flare is destroyed quickly, before it can be transported onto the night--side. Both the Flare\_UV and Flare\_Full simulations exhibit very similar behaviour. The average ozone column is quite insensitive to the inclusion of CMEs, only reducing the average ozone column by a small amount. This is attributed to a lack of additional \hox and \nox created via Reactions~\ref{Reac:proton1}--\ref{Reac:proton3}. The ionisation rate profiles used in this work (see Figure~\ref{Fig:IonisationRates}) rapidly declines below 35\,km (which has a pressure on the day--side of  $\sim100$\,Pa), and are reduced to 0 below 15\,km ($\sim7000$\,Pa). As the ionisation rate declines, the SEPs are less important and generate less \hox and \nox molecules, which results in ozone below 35\,km not being heavily affected by CMEs, and only being strongly affected by the increased UV from the flare. Figure~\ref{Fig:OzoneDUFlares} demonstrates that the ozone column is perturbed from a non--flaring state, and that the inclusion of CMEs does not produce a significant change in the global amount of ozone. A $10^{34}$\,erg flare with an accompanied CME occurred on day 60 of the simulation, and during the peak of this flare the global mean ozone column increased from 20\,DU to 45\,DU. 

\begin{figure}
    \centering
    \includegraphics[width=\columnwidth]{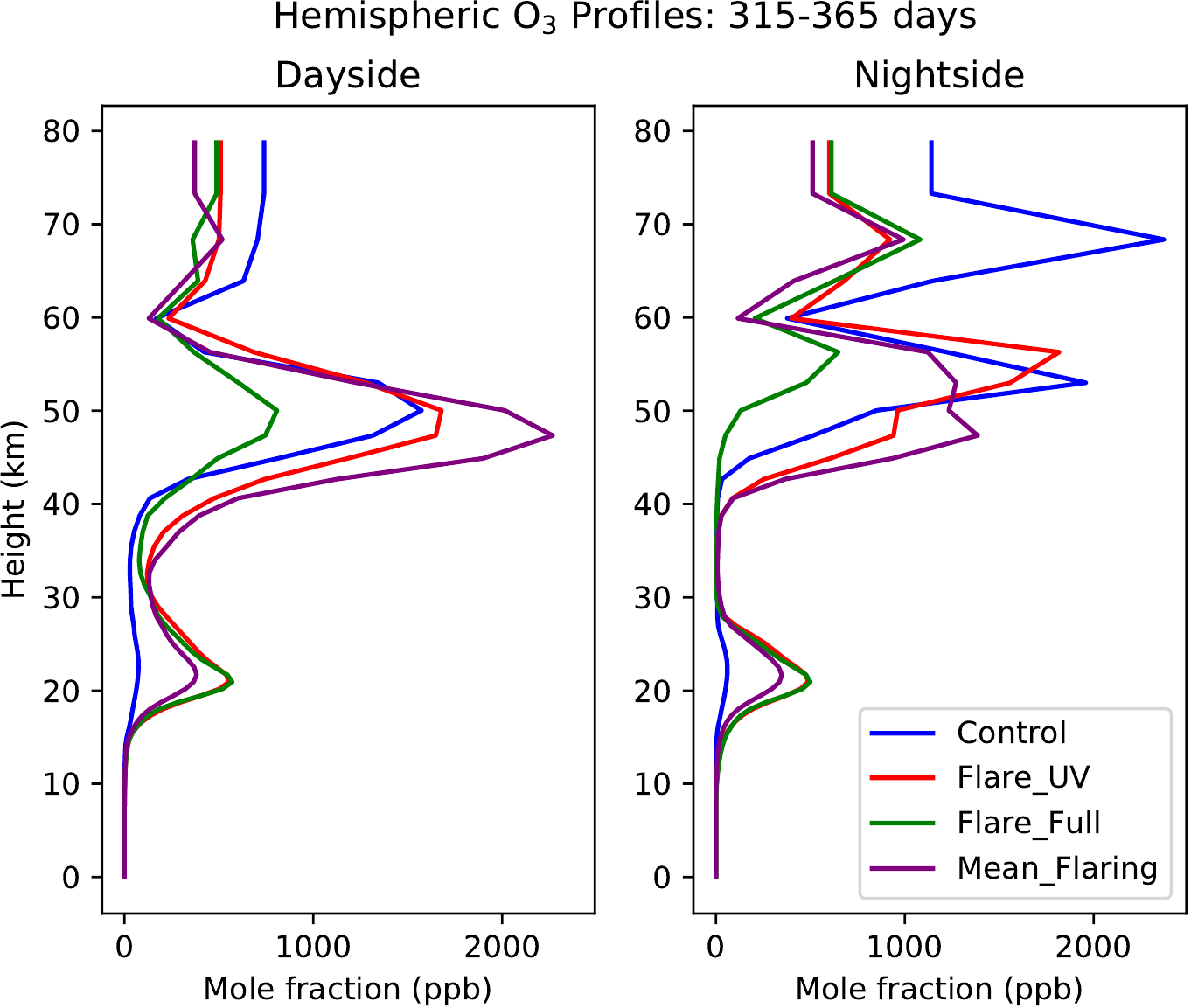}
    \caption{The spatially averaged vertical profile of the ozone mole fractions from the planet's day-side and night-side for the Control, Flare\_UV, Flare\_Full, and Mean\_Flaring simulations, averaged over the last 50 days of the simulations. The impacts of the stellar flares are seen in the generation of an ozone layer around 20-25\,km. The impact of the CMEs is seen in the depletion of ozone above 35\,km.}
    \label{Fig:VerticalOzoneProfileFlaring}
\end{figure}

Figure~\ref{Fig:VerticalOzoneProfileFlaring} shows the spatially averaged day and night--side mole fractions of ozone as a function of altitude for the Control, Flare\_UV and Flare\_Full simulations, temporally averaged over the last 50 days of the simulation. The impact of the flares, and also the SEPs is clear. An ozone layer between 20-25\,km (hereafter referred to as the lower ozone layer) has developed. This layer is also present in the quiescent simulation but is much smaller. The ozone layer between 45-55\,km (hereafter referred to as the upper ozone layer) from the Flare\_Full simulation has been depleted relative to the Flare\_UV simulation. The depletion is due to the increased amount of \ce{NO} which was generated by the CMEs. The impact of the CMEs on ozone concentration is found to have negligible long-term effects below 35\,km, as the SEP-induced ionisation rapidly declines in strength below 35\,km and generates less \hox and \nox at those altitudes as a result.

\begin{figure*}
    \centering
    \includegraphics[width=0.85\textwidth]{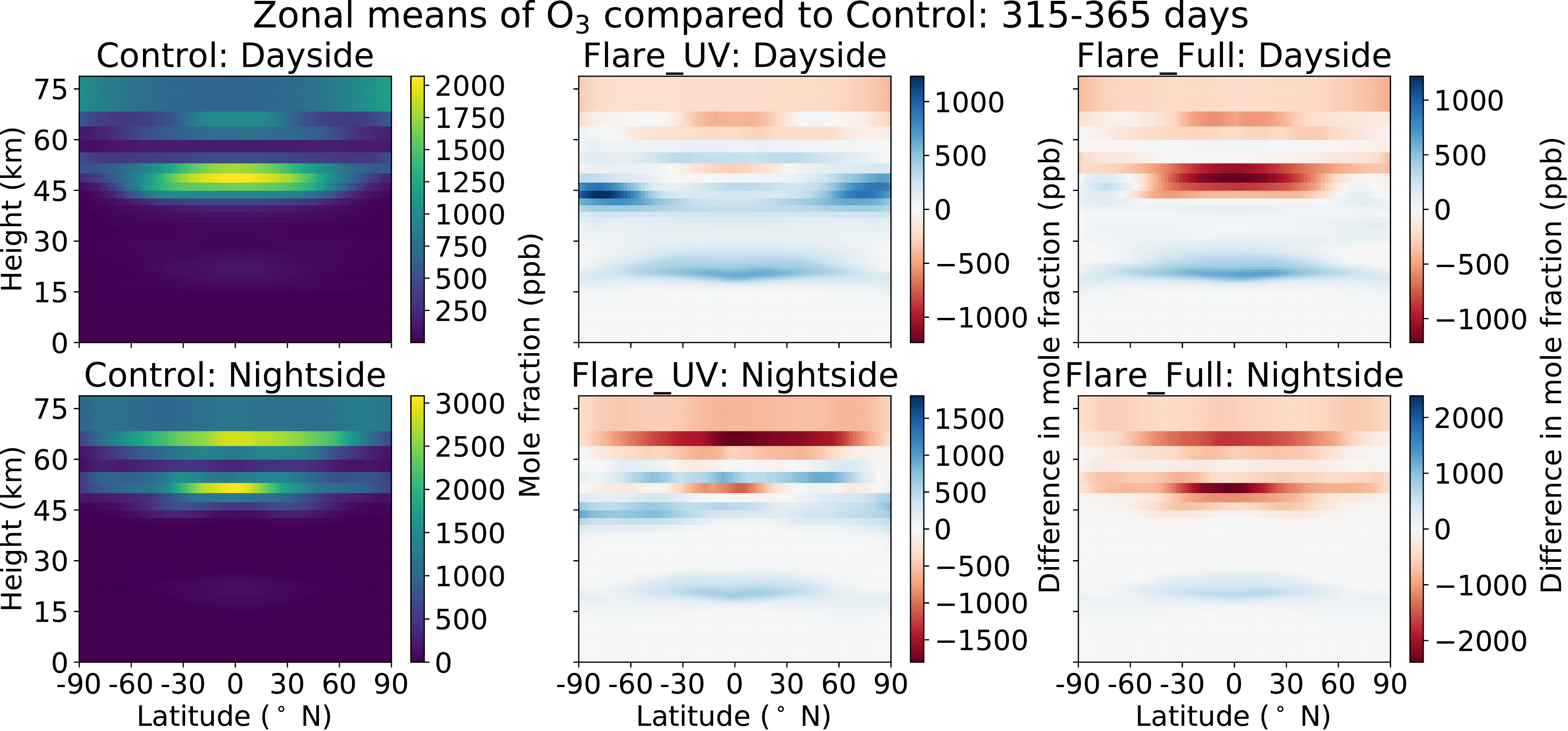} \\ \vspace{0.205cm} 
    \includegraphics[width=0.85\textwidth]{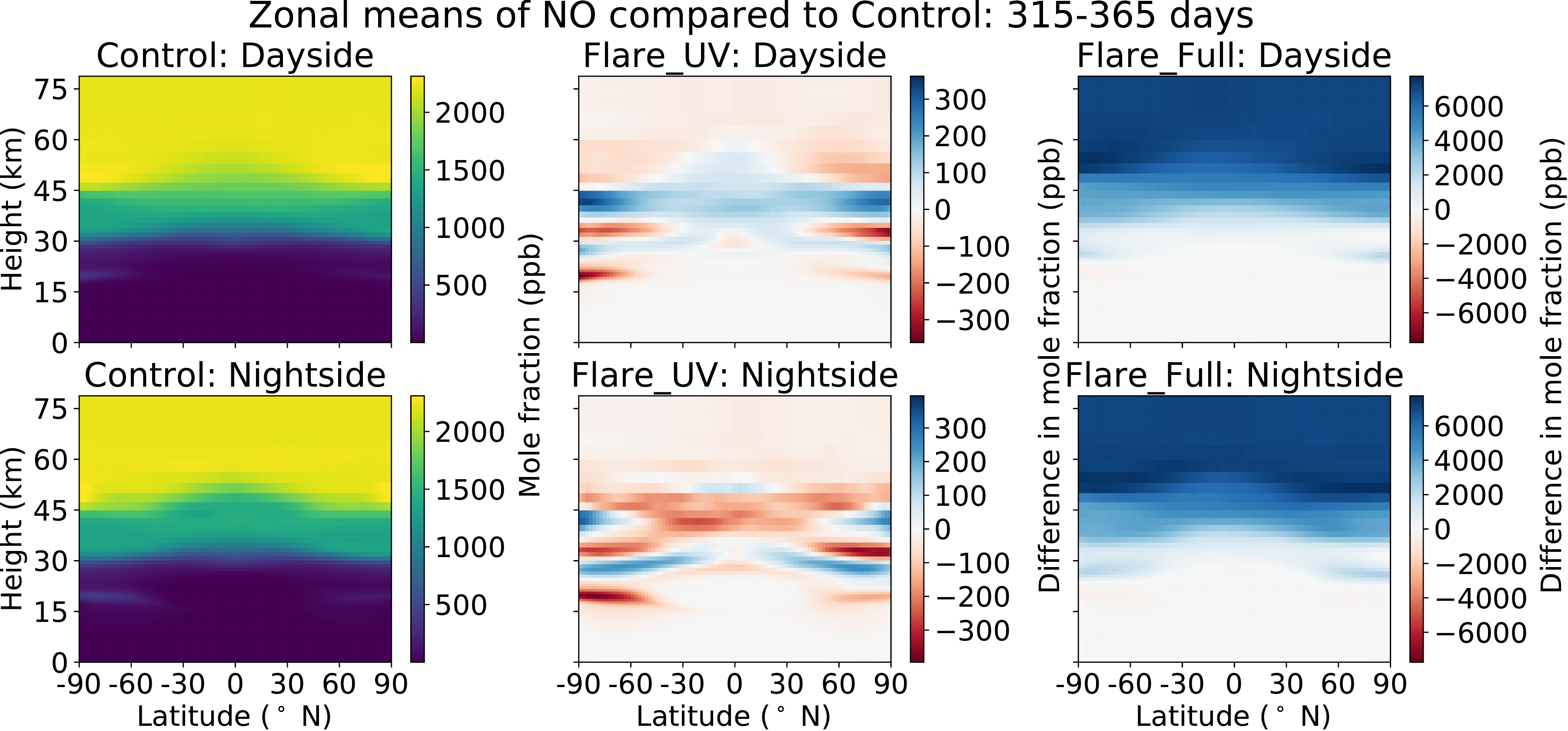} \\ \vspace{0.205cm}
    \includegraphics[width=0.85\textwidth]{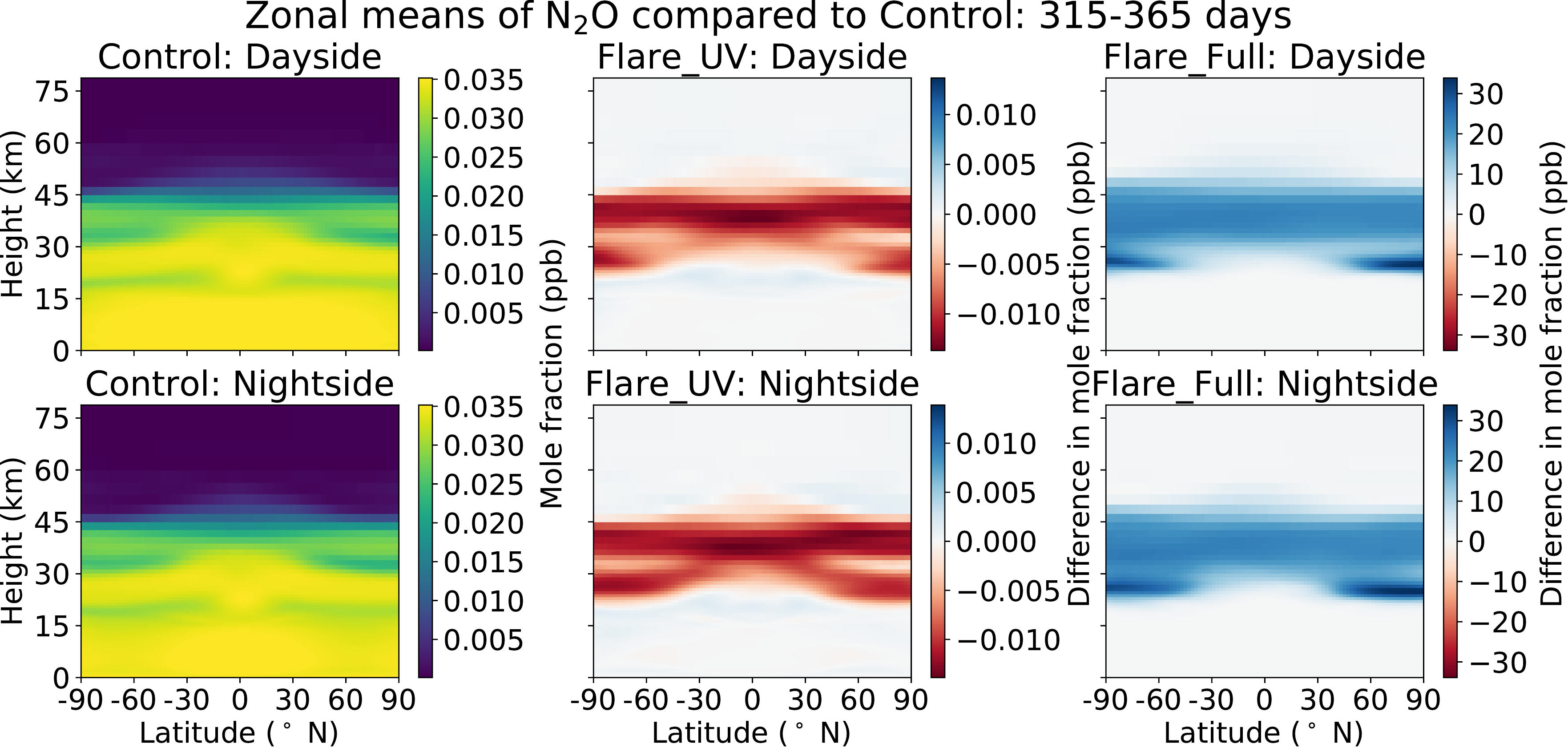}
    \caption{Zonal means of the ozone (top), \ce{NO} (centre), and \ce{N2O} (bottom) mole fraction on the day-side (top row for each molecule) and the night-side (bottom row for each molecule) of the planet for the quiescent Control (left), and differences in species mole fraction from the control from Flare\_UV (centre) and Flare\_Full (right), temporally averaged over the last 50 days of the runs.}
    \label{Fig:AllZonal}
\end{figure*}

\begin{figure*}
    \centering
    \includegraphics[width=\textwidth]{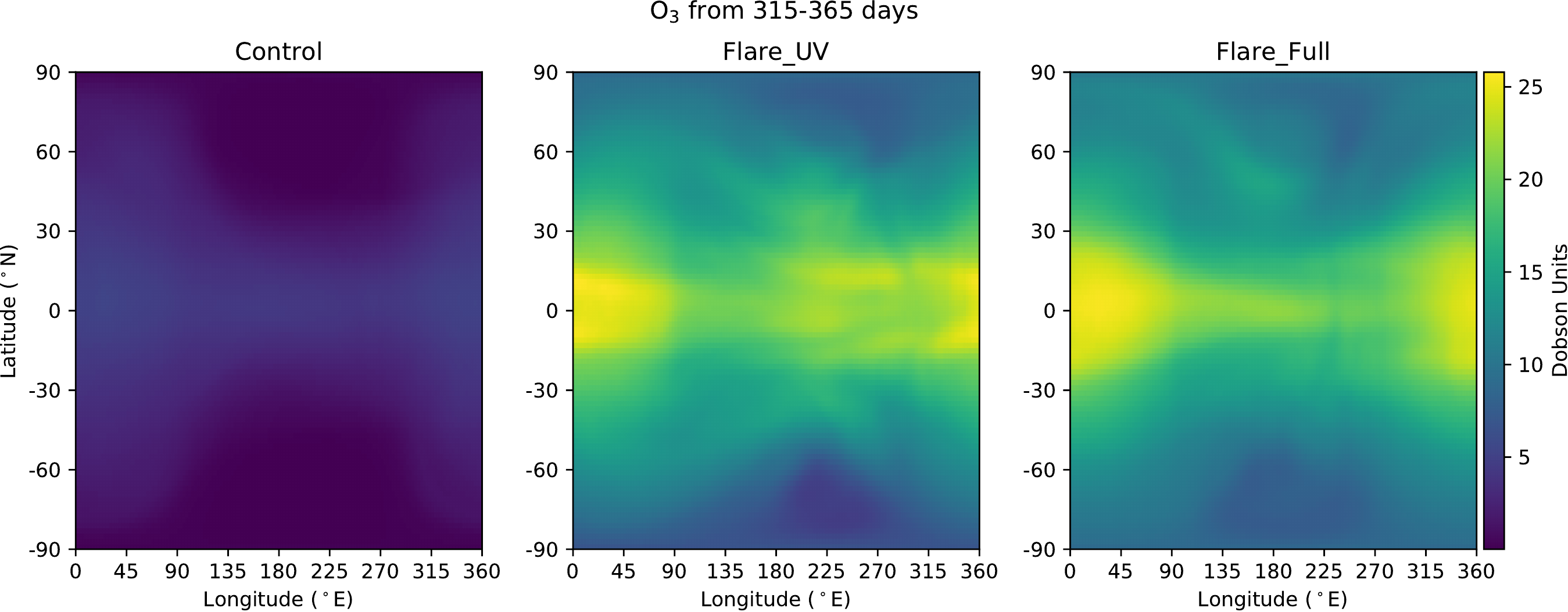}
    \caption{The spatial distribution of the ozone column (DU) for Control, Flare\_UV, and Flare\_Full averaged over the last 50\,days of the simulations. The addition of flares have significantly increased the amount of ozone in the atmosphere. Ozone is concentrated in the equatorial regions and has a larger presence on the day-side.}
    \label{Fig:OzoneCompLinear}
\end{figure*}

Figure~\ref{Fig:AllZonal}~(top) shows the zonal (east-west) average of ozone mole fraction on the planets day-side and night-side, temporally averaged over the last 50 days of the simulation period as compared to the control simulation without flares (left). The impact of flares is readily seen in the enhancement of ozone around the equatorial jet between 20-25\,km (the lower ozone layer), CMEs were found to have negligible effects at this altitude. In the upper ozone layer between 45-55\,km, we see that flares extend the upper ozone layer into the polar regions. In contrast, we see that the CMEs deplete the upper ozone layer. This is also clear in the spatial distribution of the ozone column shown in Figure~\ref{Fig:OzoneCompLinear} (which is also temporally averaged over the last 50 days), for the flaring and control simulations. Figure~\ref{Fig:OzoneCompLinear} (centre and right columns) shows that the enhanced ozone column is largest in the tropics, due to oxygen photolysis being strongest in the sub-stellar region and the ozone carried in the equatorial jet.

The results from this work differ significantly from the 1D simulations performed by \citet{Tilley2019}. They found that ozone only showed a small depletion in response to the electromagnetic portion of flares, and a much heavier depletion once the effects of CMEs were included. As in our work, they assume that every flare has an associated CME and account for the impact probability. There are likely differences in the results that are associated with adopting a 1D or 3D model. Their 1D model is a day--night average, and does not include any transport. Our results show enhanced ozone transport to the night--side, which is not captured in a 1D model. The storage of ozone in the night--side portion of the equatorial jet assists in the creation of an enhanced ozone layer. There are other differences with the components of the model. Our work uses ProxCen for the host star, whereas they used Ad Leonis for their host star. \citet{Tilley2019} also directly inject \ce{NO} and \ce{NO2} into the atmosphere instead of through the creation of atomic nitrogen, which may cause changes in several reactions. A full comparison of our stellar spectra, chemical networks (and reaction rate coefficients), photolysis calculations, and the modelling of flares and CMEs would be required to properly determine why we find that flares lead to an increase in the amount of ozone while \citet{Tilley2019} have the opposite conclusion.

The flares also induce long--term changes in the concentrations of several N--containing species. The SEPs due to CMEs create a large enhancement of \ce{NO} and \ce{N2O} in the stratosphere. For \ce{NO}, Figure~\ref{Fig:AllZonal}~(middle) shows that \ce{NO} responds to flares and CMEs very differently at different altitudes. The CMEs have lead to an increase in the amount of \ce{NO} above 30\,km by a factor of 3-4, from 1-2.3\,ppm to 9.5\,ppm. This increase is due the following reactions, with the first one being controlled by the SEPs,\begin{chemequations}
    \begin{alignat*}{3}
        \ce{N2 &-> N(^2D) + N(^2D)},\\
        \ce{N(^2D) + O2 &-> NO + O(^3P)}.
    \end{alignat*}
\end{chemequations}


Figures~\ref{Fig:AllZonal}~(bottom) show the impacts of the stellar flares on \ce{N2O}. These figures show that the UV irradiation causes minimal changes in the mole fraction of \ce{N2O}, but the SEPs have induced a very large increase between 25-50\,km. We attribute this to
\begin{chemequations}
    \begin{alignat*}{3}
        \ce{N2 &-> N(^4S) + N(^4S)},\\
        \ce{NO + O3 &-> NO2 + O2}, \\ 
        \ce{N(^4S) + NO2 &-> N2O + O(^3P)},
    \end{alignat*}
\end{chemequations}\\
as the SEPs cause significantly more \ce{N(^4S)} and \ce{NO2} (via the creation of additional \ce{NO}) to be generated, which would enhance this reaction and generate more \ce{N2O}. 

\begin{figure*}
    \centering
    \includegraphics[width=\textwidth]{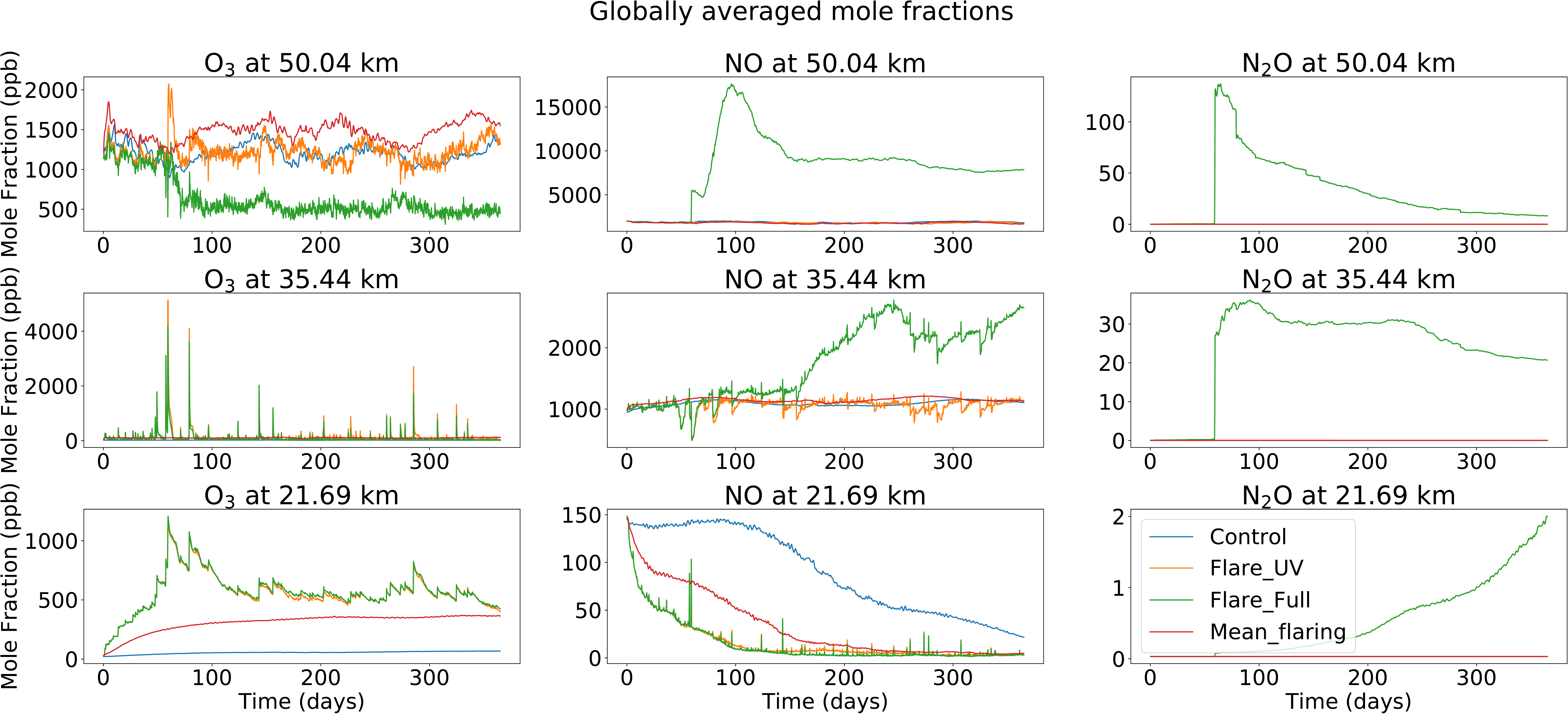} 
    \caption{The globally averaged mole fraction of ozone (left column), \ce{N2O} (centre column), and \ce{NO} (right column) in the lower ozone layer (bottom row), mid stratosphere (middle row), and the upper ozone layer (top row).}
    \label{Fig:AllHeights}
\end{figure*}

Figure~\ref{Fig:AllHeights} shows the temporal evolution of the globally averaged mole fraction of ozone, \ce{NO}, and \ce{N2O} at several heights, chosen to sample three regions seen in Figure~\ref{Fig:VerticalOzoneProfileFlaring}; the upper ozone layer, the mid-stratosphere, and the lower ozone layer. We see that different regions of the atmosphere respond to flares and CMEs in very different ways in our simulations. The impact of flares and CMEs on each molecule will be discussed individually. Figure~\ref{Fig:AllHeights}~(left) shows the evolution of the globally averaged mole fraction of ozone. The lower ozone layer rapidly grows from $\sim$25\,ppb to a concentration of 400-500\,ppb which is perturbed by stellar flares. 
The abundance of ozone in the mid-stratosphere region is quite sensitive to the flares and shows rapid increases and decreases in ozone concentration. The impact of the SEPs is also quite visible, but not long lasting, as the ozone abundance rapidly decreases after the end of the flare. The abundance of ozone from the Mean\_flaring simulation remains below the values observed in other flaring simulations, but the result at the end of all flaring simulations is similar.

The upper ozone layer shows little changes (compared to Control) in concentrations due to flares (aside from the $10^{34}$\,erg flare on day 60 of the simulation which has a short--lived increase of 750\,ppb relative to Control), but does show a response to CMEs, causing the upper ozone layer to reduce in concentration from $\sim1250$\,ppb to 500\,ppb. A long--lived change in the ozone concentration is visible after the $10^{34}$\,erg flare and maximum strength CME impacts the planet on day 60 of the Flare\_Full simulation. We see that the upper ozone layer in the Flare\_Full simulation stays at a lower ozone concentration and does not recover towards the other simulations, even during periods of relatively quiet activity such as the period between days\,340-365 of the simulation, at least over the duration of these simulations. The abundance of ozone from the Mean\_flaring simulation is enhanced compared to Flare\_UV, telling us that while the short-term effects of flares do not individually change the upper ozone layer, the temporally--resolved stellar flares are important to include to accurately model the evolution of the upper ozone layer. Comparing the results from the flaring simulations to those reported by \citet{Chen2020} (Figure 3), we see qualitatively similar results for the long term trend of ozone at this altitude, where both show a long-term depletion of ozone. 

Our results are also similar to the recent work simulating the super--Earths GJ~832~c and GJ~581~c by \citet{Louca2022} who reported temporary enhancement of ozone at similar altitudes as our lower ozone layer and depletion of ozone at a similar height to our upper ozone layer.


As seen in Figure~\ref{Fig:AllHeights}~(bottom centre), in the lower stratosphere \ce{NO} experiences temporary increases due to the flares (and only minimal responses to CMEs) and rapidly returns to the concentrations it was at before the flare began. At mid altitudes, we see that \ce{NO} shows a negative response to stellar flares (as it depletes during a flare and recovers afterward) but does have a positive response to CMEs, eventually leading to a long-term enhancement in concentration. We attribute this to the creation of \ce{NO} at high altitudes, which is transported to lower altitudes. At high altitudes, we see that \ce{NO} shows little to no response to flares, but shows a very strong response to CMEs. The peak enhancement increases the concentration by a factor of 9, reaching 18 ppm before decreasing to 8-9\,ppm which is maintained for the rest of the simulation, albeit slowly decreasing. This was not solely caused by the $10^{34}$\,erg flare and CME that occurred on day 60 of the simulation (although the impact is clearly noticeable through a very rapid increase in concentration from 2-5\,ppm on the bottom figure) but by a series of weaker flares. This indicates that the largest flares are not the major cause of changes in \ce{NO}. Instead the cumulative impacts of the weaker flares and CMEs are the main driver in the changes of \ce{NO} concentration. We do see that the results from the Mean\_flaring do mostly agree with the Flare\_UV simulation.

Figure~\ref{Fig:AllHeights}~(right column) shows that the additional atomic nitrogen generated by CMEs is able to temporarily enhance \ce{N2O} concentrations up to 100\,ppb at high altitudes in our simulations. This is above the levels of \ce{N2O} seen in \citet{Segura2003} in simulations of Earth-like atmospheres including surface fluxes of \ce{N2O}. On Earth, \ce{N2O} is mostly produced by biological activity and is thought to be a biosignature \citep{Segura2003}. Our results, however, show that care must be taken when interpreting enhanced \ce{N2O} as an indicator of biotic processes. Comparing our results to \citet{Chen2020}, we find that our results differ significantly. \citet{Chen2020} report \ce{N2O} abundances significantly higher than found in our results, with a peak in \ce{N2O} mixing ratio of $10^4$\,ppb during their flare peak, almost two orders of magnitude higher than our results. This, in part, may well be due to the pre--flare conditions also having a significantly higher abundance than in our case, and should be investigated in future work, beyond the scope of this initial study (see discussion in Section~\ref{subsec:future_work}). Overall, however, our results and those of \citet{Chen2020} exhibit similar qualitative behaviour.

\subsubsection{Planetary Habitability}
\label{Sec:PlanetHab}
The surface UV radiation environment is a useful gauge of the impact that flares may have on a planet's habitability. However, the lack of wet deposition in this work creates a large concentration of \ce{HNO3} (approximately 60\,ppm) throughout the troposphere, which acts as a strong UV absorber, heavily impacting the surface UV radiation environment. The contribution of \ce{HNO3} to photoabsorption was removed in order to crudely approximate the inclusion of wet deposition. A short test was conducted to determine how this removal of \ce{HNO3} would affect the atmospheric composition. While the details of the atmospheric composition do change quantitatively to some extent, the bulk composition does not, and the qualitative result of this work (flares generating ozone and CMEs having a limited impact on atmospheric composition) will remain unaffected. For completeness, versions of the Figures described in this section and Section~\ref{Sec:PlanetObs} with the contribution of \ce{HNO3} are included in Appendix~\ref{appsec:withHNO3}. In future work we plan to implement a proper wet deposition scheme but this requires development and exploration of the correct choices of controlling parameters.

\begin{figure*}
    \centering
    \includegraphics[width=\textwidth]{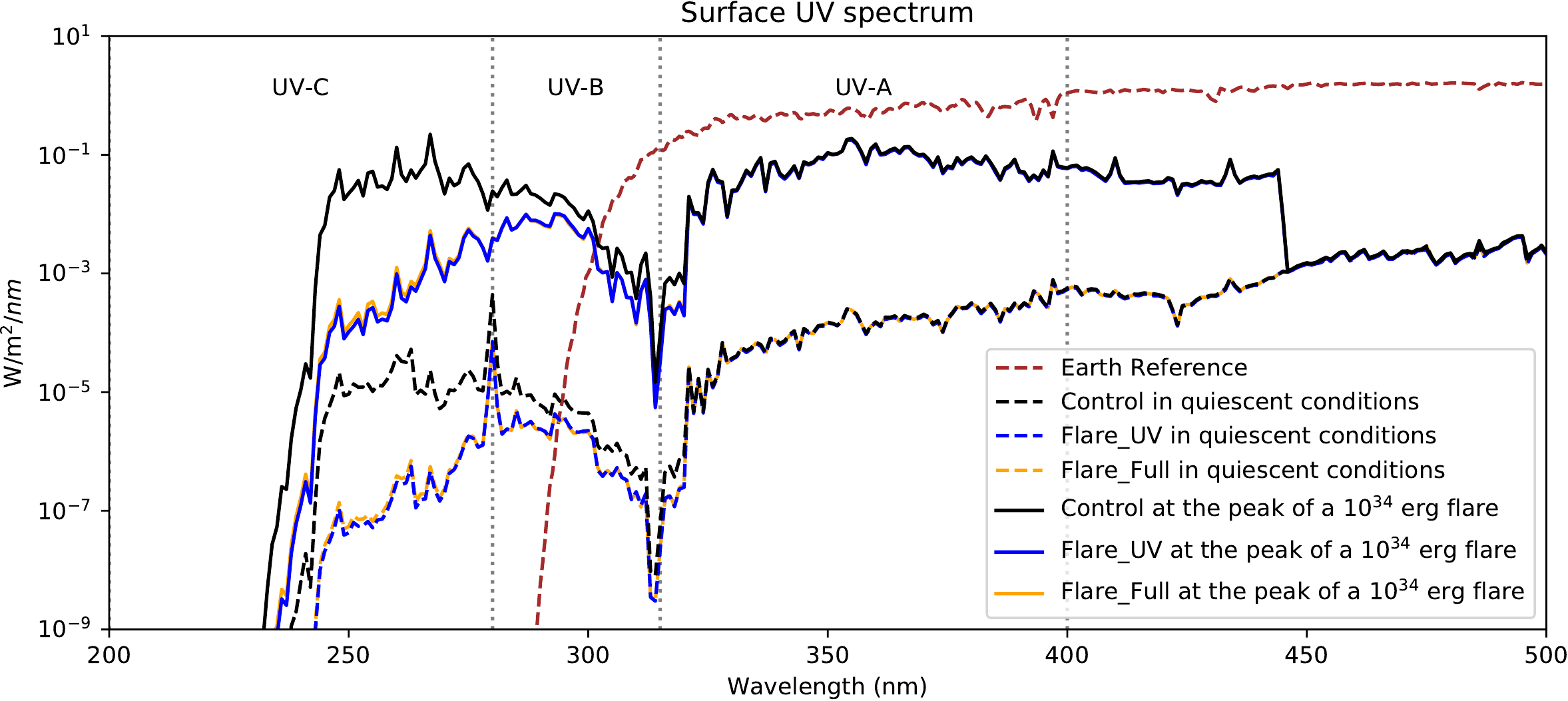}
    \caption{The average day-side surface UV environment under quiescent conditions and during the peak of a $10^{34}$\,erg flare from the end of the spin-up simulation Quiet\_Full (black), a run containing only stellar flares (blue), and a simulation containing both stellar flares and CMEs (orange), using the atmospheric configurations from the end of their respective simulations. A reference for the Earth under quiescent conditions is also included. The changes in atmospheric composition due to stellar flares have caused additional screening of the surface from UV radiation.}
    \label{Fig:UV}
\end{figure*}

Figure~\ref{Fig:UV} shows the average day-side surface radiation environment for the Control, Flare\_UV, and Flare\_Full simulations under quiescent conditions (dashed) and at the peak of a 10$^{34}$\,erg flare (solid). The Control simulation is only run with quiescent conditions, but seeing what the surface UV environment in Control would have been if it was subjected to a strong flare is useful as a comparison to simulations which have already been subjected to many flares and CMEs. A reference spectrum for the Earth under quiescent conditions (American Society for Testing and Materials G-173-03 reference spectra\footnote{\url{https://www.astm.org/g0173-03.html}, spectra obtained from \\ \url{https://www.nrel.gov/grid/solar-resource/spectra-am1.5.html}}) is included as well, demonstrating that our simulations result in a significantly different surface UV environment.

Comparing Control under quiescent conditions to the quiescent Earth reference, we see that the planet receives significantly less UV--A and UV--B, but much more UV--C. The changes to Flare\_UV and Flare\_Full are apparent even during quiescent conditions. The changes in atmospheric composition due to stellar flares reduce the amount of UV radiation below $320$\,nm that reaches the planets surface. The UV--A flux is relatively unaffected, but UV--B and UV--C have been significantly reduced. This occurs for quiescent conditions and during the flare peak, the changes in atmospheric composition have added additional screening of UV which has resulted in a relatively modest decrease in UV--A (315--400\,nm) and UV--B (280--315\,nm), and a much larger reduction in UV--C (200--280\,nm).  Given the dramatic increase in the flux emitted by the star at these wavelengths during flares that can reach the surface, as seen in the Control simulation at the flare peak, this implies a stabilising feedback through the generation of a `shielding' layer. This could have important implications for the existence of life on such planets. Interestingly, we see that the Flare\_UV simulation shows a greater decrease in the surface UV than that found in the Flare\_Full case. While this is most noticeable at shorter wavelengths where the surface fluxes are very low, this tells us that the species created by the CMEs cause an overall reduction in the amount of UV shielding. The amount of UV--B and UV--C which reaches the planet's surface during a flare are still much higher than that seen on Earth however, which has implications for habitability which are discussed in Section~\ref{Sec:PlanetHab}.

\begin{figure*}
    \centering
    \includegraphics[width=\textwidth]{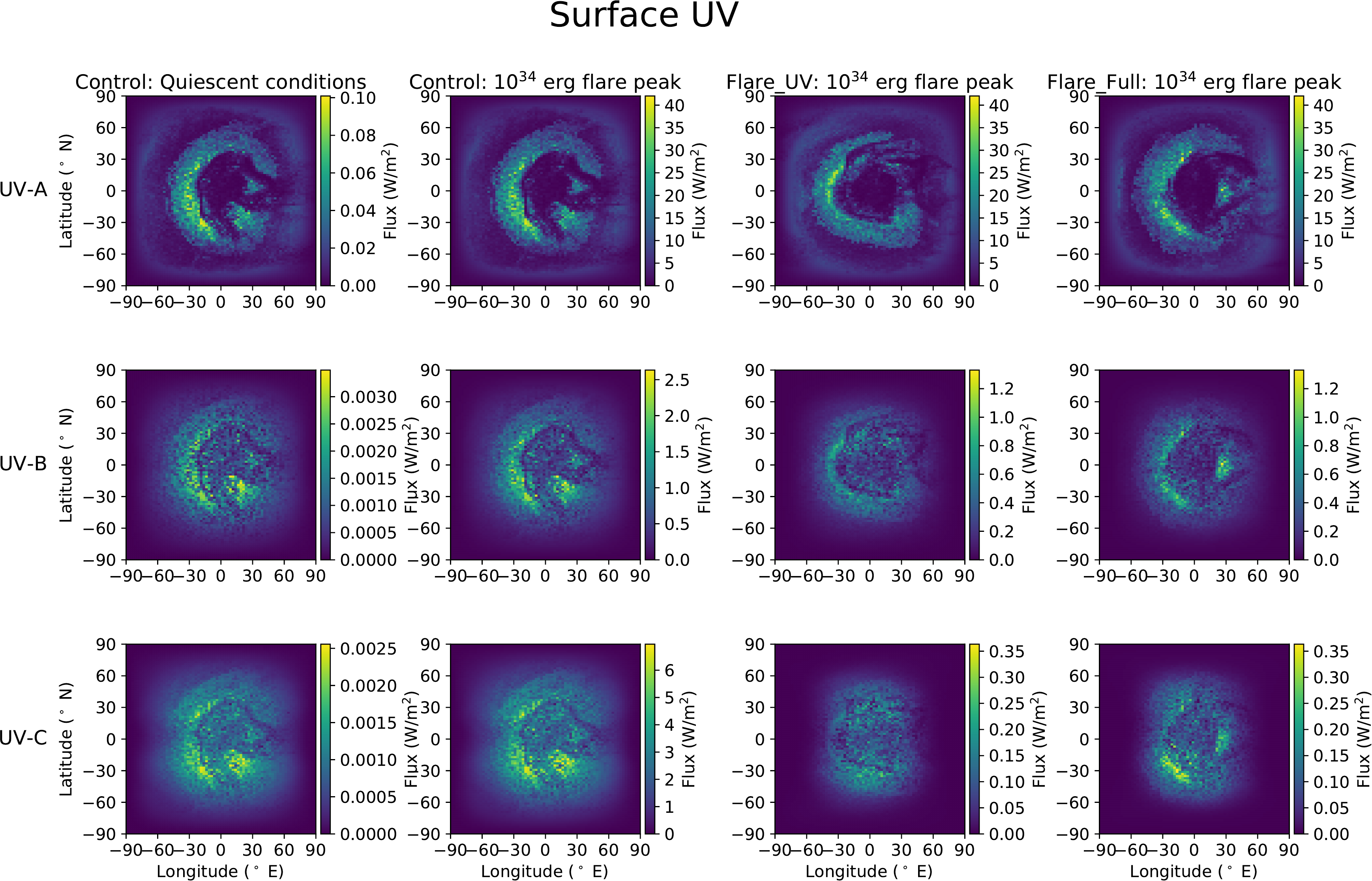} 
    \caption{The surface UV radiation environment for the Control simulation under quiescent conditions, as well as Control, Flare\_UV, and Flare\_Full if they were to be subject to the peak of a $10^{34}$\,erg flare. As seen previously in Figure~\ref{Fig:UV}, the changes in atmospheric composition due to flares have not significantly changed the UV--A flux, but have resulted in a a significant reduction in the amount of UV--B and UV--C flux.}
    \label{Fig:QuiUV}
\end{figure*}

It should be noted that our simulations of ProxCen~b have a very cloudy region around the substellar point which will reduce the average day-side surface UV drastically. Cloud formation is most prominent on the day--side around the substellar point. This is due to our simulation containing a global ocean, the presence of an ocean at the substellar point drives cloud formation, \citep[see][for more details]{Sergeev2020}. Figure~\ref{Fig:QuiUV} shows the spatial distribution of the surface UV flux on the day--side of the planet for three cases; Control, Flare\_UV, and Flare\_Full. This is separated into the UV--A (315-400\,nm), UV--B (280--315\,nm), and UV--C (200--280\,nm) bands. As shown in Figure~\ref{Fig:UV}, the surface UV--A flux is not meaningfully altered by the presence of flares or SEPs. There is, however, a 50\% reduction in the maximum surface UV--B flux from a peak of 2.6\,W\,m$^{-2}$ to approximately 1.3\,W\,m$^{-2}$, and a reduction in the maximum UV--C flux from 6.95\,W\,m$^{-2}$ to 0.36\,W\,m$^{-2}$, a 94.77\% reduction, both of which we attribute to the increased amount of ozone. The region around the substellar point receives less UV than the areas further away from the substellar point, suggesting that this region may be affected less by stellar flares due to the large amounts of cloud. A planet with a different land--ocean configuration may behave differently in this regard as changes in the hydrological cycle from a different configuration will affect the generation of clouds \citep{Lewis2018},  A planet without a large source of water (an ocean, sea, or series of lakes/swamps) in the warmer regions of the planet (in our planets case the substellar point) will have significantly less cloud formation. Fewer clouds would mean that the amount of UV which reaches the surface would increase. Likewise, if the planet was warmer or colder (due to being closer or further from its star, or the star itself being hotter or colder), the regions of the planet where clouds could form would also change. As well, changes in the land-ocean configuration and hydrological cycle will result in changes to dry deposition and wet deposition rates, further changing the atmospheric composition.

\begin{figure*}
    \centering
    \includegraphics[width=\textwidth]{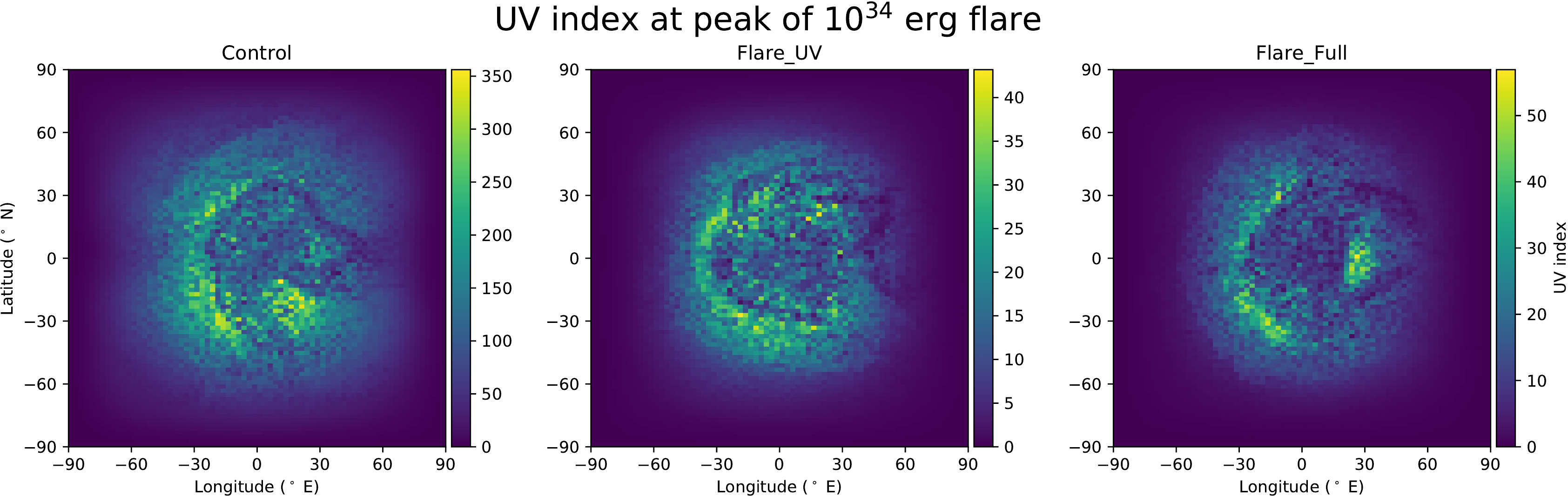}
    \caption{Maps of the UV index for Control, Flare\_UV, and Flare\_Full if they were to be subject to the peak of a $10^{34}$\,erg flare. While the UV index is extremely high for all simulations, changes in atmospheric composition due to previous flares result in the peak UV index being reduced by $\sim$85\%.}
    \label{Fig:UVIndex}
\end{figure*}

The high UV flux during a flare presents a danger for any life which may exist on the planets surface. As a proxy for examining the effects of flares on surface life, we can use the UV index. The UV index is used to measure the danger of sustained exposure to sunlight to human skin. It is calculated as 
\begin{equation}
    \text{I}_{UV} = k_{er} \int_{250\,\text{nm}}^{400\,\text{nm}} E(\lambda)S_{er}(\lambda) d\lambda,
\end{equation}
where $E(\lambda)$ is the solar spectral irradiance, $S_{er}(\lambda)$ is the erythema action spectrum, which represents the relative effectiveness of UV radiation to damage human skin, and $k_{er}$ is a constant of $40$\,m$^2$W$^{-1}$ which was chosen so the standard range of the index is between 0 and 10 (the range of values typically seen on Earth). An idealised form of the erythema action spectrum (International Organization for Standardization standard ISO 17166:1999\footnote{\url{https://www.iso.org/obp/ui/\#iso:std:iso:17166:ed-1:v2}}) is described as
\begin{align}
S_{er}(\lambda) =   \begin{cases} 
                        1   &250 < \lambda < 298\,\text{nm}, \\
                        10^{0.094(298-\lambda)} &298 < \lambda < 328\,\text{nm}, \\
                        10^{0.015(139-\lambda)}  &328 < \lambda < 400\,\text{nm}.
                    \end{cases}    
\end{align}
During quiescent conditions the UV index is very mild (less than 0.2 for the Control simulation, and less then 0.04 for Flare\_UV and Flare\_Full, due to the increased shielding from ozone as previously described), but the UV index reaches very high values during the peak of a maximum strength flare, as shown in Figure~\ref{Fig:UVIndex}. The Control simulation has a peak UV index of over 350. The ozone generated by previous flares has reduced the UV index to $\sim$55, which is still extremely high, but is a reduction of $\sim$85\% from Control. Using Equation~\ref{Eqn:InvCumSum}, flares of this magnitude occur every 500 days on average. This presents a key danger for the surface habitability of this planet, even after a shielding layer of ozone has been generated. Future research on whether life could adapt to these conditions should be conducted.
\subsubsection{Potential Observability}
\label{Sec:PlanetObs}
To determine whether the impacts of stellar flares could be observable for our representation of ProxCen~b, we generated a synthetic transmission spectra for several simulations. The UM is able to output a synthetic transmission spectrum, using the method described in \citet{Lines2018b} and recently updated by \citet{Christie2021}. It should be noted that ProxCen~b is not thought to transit \citep{Jenkins2019}. The planet we simulate is merely a planet based on the parameters of ProxCen~b, with the results indicative for M dwarf hosted planets with `Earth--like' atmospheres. As stated in Section~\ref{Sec:PlanetHab}, we have removed the contribution of \ce{HNO3} to the transmission spectrum. See Appendix~\ref{appsec:withHNO3} for a discussion of the results with the contribution of \ce{HNO3}.

\begin{figure*}
    \centering
    \includegraphics[width=\textwidth]{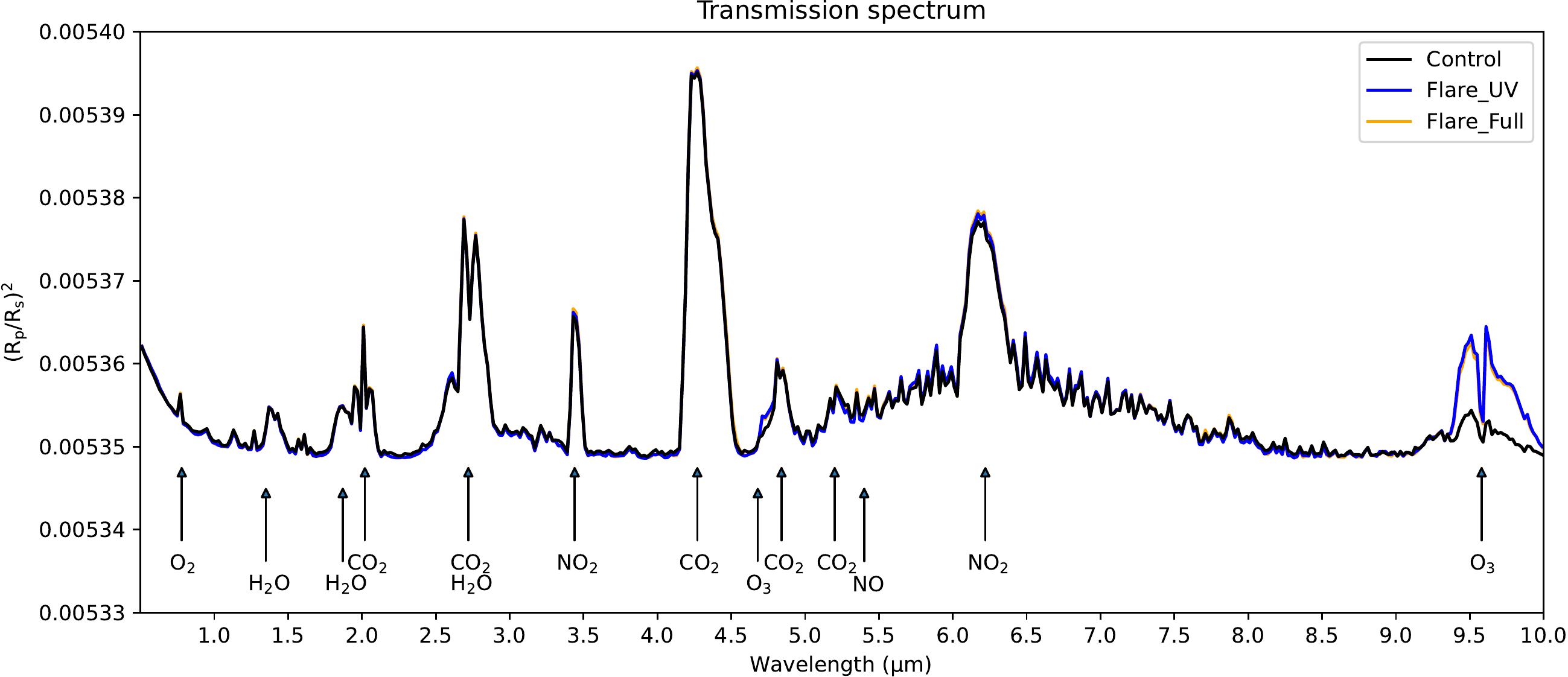}
    \caption{The transmission spectra for the simulated planets for 500\,nm-10\,$\mu$m at the end of their respective simulations. The colors refer to the same simulations as Figure~\ref{Fig:UV}.}
    \label{Fig:Transmission}
\end{figure*}

Figure~\ref{Fig:Transmission} shows the transmission spectra (between 500\,nm and 10\,$\mu$m) from the previously described simulations (Control, Flare\_UV, and Flare\_Full). Changes in the transmission spectrum are caused by changes in atmospheric composition, temperature, and pressure. Figure~\ref{Fig:Transmission} shows that the transmission spectrum is rather unchanged by the changes in the atmospheric composition due to flares or SEPs. We observe strong absorption features for \ce{NO2} and ozone. The ozone absorption peaks at 9.5\,microns are the only features altered in a noticeable fashion. The differences in these features are relatively small and are not expected to be readily discernible with current-generation instrumentation. 

The atmospheric composition changes due to the SEPs (a reduction of the upper ozone layer, and a significant increase in \ce{NO} and \ce{N2O}) cause very small changes in the transmission spectrum, namely a slight increase in the continuum absorption in several wavelength regions, and a small increase in the \ce{NO2} absorption feature at 6.3\,microns.


\section{Conclusions}
\label{sec:conclusions}
In this work we have coupled the UM, a 3D general circulation model, a chemical kinetics scheme, and a photolysis scheme to create a self-consistent photochemical scheme capable of describing the interplay between atmospheric chemistry and planetary dynamics. 

In this first application of our model, we simulated Proxima Centauri~b as an indicative terrestrial aquaplanet, although our qualitative results should hold for similar targets. The simulated planet is tidally locked with an Earth-like atmosphere orbiting an M dwarf star. We incorporate Earth-based observations of ionisation caused by stellar protons to create an approximation of a quiescent stellar wind, as well as a representation of the ionisation caused by a CME which is used to approximate the effects of CMEs which result in SEPs impacting the planet's atmosphere through the creation of short-lived radical species, which induce significant changes in the atmosphere. We find that under quiescent conditions the planet maintains an extremely thin upper ozone layer at 45-55\,km resulting in an average ozone column which is hundreds of times thinner than seen on Earth. We find that the introduction of \hox and \nox chemistry results in the depletion of ozone globally as expected, but the depletion is strongest through the inclusion of \nox chemistry. The depletion of ozone due to the introduction of \nox chemistry is strongest within the night--side gyres, removing a night--side reservoir of ozone.

We find that the effect of stellar flares on the planet's atmosphere is to increase in the amount of ozone present in the atmosphere. A lower ozone layer is developed between 20-25\,km and is carried to the night-side through the equatorial jet. SEPs are found to cause minimal changes in the total amount of ozone, but do cause a reduction in the upper ozone layer. 

The increased UV radiation due to the stellar flares causes a 20\,DU increase in the amount of ozone present. The ozone concentration enters a punctuated equilibrium state which is perturbed by flares to temporarily increase the ozone concentration. CMEs were found to have a limited impact on the amount of ozone. The impact is dependent on altitude. At low altitudes the ozone concentration increases from $\sim10$\,ppb to hundreds of ppb. At mid altitudes we see a rapid increase and decrease in ozone concentration, with the increases ranging from hundreds to thousands of ppb. At high altitudes we do not see a response due to flaring. We see a response due to CMEs which reduces the ozone concentrations by $\sim600$\,ppb. This reduction indicates a long-term divergence in the chemical evolution of ozone in the upper atmosphere due to CMEs.

Simulating the planet with a constant spectrum consisting of the time-averaged stellar spectra from the year-long simulation shows a mixture of agreement and disagreement with the simulation only containing flares. While the concentrations of species such as \ce{N2O} and \ce{NO} broadly match, species that are more sensitive to flares (such as \ce{O3}) are quite different.  This tells us that while such a spectrum can be used to model changes in atmospheric composition due to stellar flares without needing a high-resolution time-varying stellar irradiance model, it cannot be used without caution, as it is not accurate for every species and will not capture any short-term behaviours.

The introduction of CMEs causes a significant increase in the concentration of the biosignature \ce{N2O} in the stratosphere. While this increase is not readily discernible by the current generation of instrumentation, it does highlight the need for caution if \ce{N2O} is detected in significant quantities, as we have found that stellar flares and CMEs can plausibly act as an abiotic source of \ce{N2O}.

The changes in the atmospheric composition have heavily reduced the amount of UV--B and UV--C which reaches the planets surface. While rare, during the peak of the strongest flares in this model, the surface UV--A and UV--B fluxes increase by a factor of 400 and the UV--C by a factor of 145. These results were obtained after removing the contribution of \ce{HNO3} to radiative absorption, mimicking perfect wet deposition.

To summarise, we find that the stellar flares are inducing changes in the atmosphere to create additional shielding from UV. Essentially, the atmosphere is responding in such a way that makes the next flare less impactful, with the most significant, and potentially irreversible impacts being caused by the largest flares which are relatively common for M dwarf stars.

\subsection{Future work}
\label{subsec:future_work}
The question of whether M dwarf hosted planets are habitable is a challenging one. In general, for 3D exoplanet climate modelling studies can adapt complex and more `complete' treatments developed for Earth, or develop more simplified treatments. For the former case the studies are less likely to miss important elements, and could be more accurate, however the latter approach is more amenable to interpretation and less subject to Earth--centric assumptions. In reality, to make progress we need a range of approaches. In this work we introduce a new model, complementary in its approach to that of \citet{Chen2020} and \citet{Braam2022} in examining ozone chemistry in 3D, but are aware that this first step motivates extensive follow--up both in terms of more extensive application of this model, and further development.

Firstly, for our current study, improvements in computational efficiency would allow us to perform longer simulations including a larger number of flares, and studying the longer term behaviour of the atmosphere. The evolution of the ozone distribution is not completely understood and initial conditions for a flaring simulation are not well constrained. Work is underway to adapt the next generation climate model of the Met Office, termed LFRic \citep{Adams2019} to exoplanets (Sergeev et al., in prep), which will open up much larger scale computations allowing the inclusion of more physical process, improvement to higher spatial and temporal resolution and longer simulation times. In this work we have focused on a few important species, but as detailed in Appendix \ref{appsubsec:list_track} we track the abundances of many more species which we could explore in more detail. Additionally, we have assumed an unmagnetised planet, but could implement a spatial dependence of the SEP impacts to mimic a magnetic field topology. We have also assumed an `Earth--like' atmospheric composition for this initial study, but varying compositions could also be studied through adaptations of the chemistry. Additionally, it is clear that `M dwarfs' are not a uniform population, but actually a diverse set of objects so studies should be expanded to cover the spectral range more completely. 

For our current model, the treatment of dry deposition should be improved beyond the simple model of \citet{Giannakopoulos1999} and wet deposition added. This would act to alter the abundance of the species listed in Section~\ref{subsec:deposition} including ozone. The inclusion of wet deposition would impact the chemistry in the troposphere, and heavily deplete \nox reservoirs such as \ce{HNO3}. Additionally, non--LTE effects in the upper atmosphere should be included to more accurately capture the high altitude heating, something which is essentially omitted in this work. SOCRATES is currently being upgraded to include such a treatment \citep{Jackson2020}. Furthermore, ions and aqueous interactions could be included in the chemistry, which would lead to a more complete description of the impact of the stellar activity. The generation of hazes in the upper atmosphere would also likely impact the UV budget so should also be included. Hazes are likely to play a key role in, for example, the Archean Earth and analogue exoplanets \citep{Arney2016}. Finally, additional sources and sinks into the atmosphere could be included such as atmospheric escape and influx from the surface (e.g., volcanic activity). 

The long--term changes in the abundances in several species as a result of the flares or SEPs suggests that the occurrence and timing of large flares or CMEs might be quite important in determining the state of the planet at the end of the simulations. This could be tested in several ways, such as increasing the length of the simulation to eliminate short-term effects, or by subjecting the planet to the same sampling of flares, but in a different order.

This model could also be adapted to the study of the Archean Earth, as mentioned with the inclusion of hazes, as well as being applied to hot Jupiters with adaptations to the chemistry and model setup \citep[as the UM is already routinely applied to hot Jupiters to study the chemistry, e.g,][\& Zamyatina et al., submitted]{Drummond2020}. In particular, we plan to adapt this model to the Archean Earth, and study exoplanet analogues of this stage in Earth's history, where the first evidence of life is found \citep{Nisbet2001}.

Clearly there is much to be done to improve our understanding of the interaction between `active' stars and the planets they host. 

\section*{Acknowledgements}

RR is funded through a University of Exeter, College of Engineering, Mathematics and Physical Sciences PhD scholarship. Material produced using Met Office Software. We acknowledge use of the Monsoon2 system, a collaborative facility supplied under the Joint Weather and Climate Research Programme, a strategic partnership between the Met Office and the Natural Environment Research Council. This work was partly supported by a Science and Technology Facilities Council Consolidated Grant [ST/R000395/1], the Leverhulme Trust through a research project grant [RPG-2020-82] and a UKRI Future Leaders Fellowship [grant number MR/T040866/1]. PIP acknowledges funding from the STFC consolidator grant \#ST/V000594/1. MB acknowledges funding from the European Union H2020-MSCA-ITN-2019 under grant agreement no. 860470 (CHAMELEON). For the purpose of open access, the author(s) has applied a Creative Commons Attribution (CC BY) licence to any Author Accepted Manuscript version arising. We thank the reviewer for a thorough review of our paper.

\section*{Data Availability}


The research data supporting this publication are openly available from the University of Exeter's institutional repository at: \url{https://doi.org/10.24378/exe.4244}



\bibliographystyle{mnras}
\bibliography{references,references_extra} 




\appendix
\section{Chemical Network}
\label{appsec:reactions}
In this appendix section, we first list the 22 chemical species that we actively track in our model (Section~\ref{appsubsec:list_track}), followed by the reactions included in our chemistry and radiative transfer schemes (Section~\ref{appsubsec:reactions}). 

\subsection{Chemical Species}
\label{appsubsec:list_track}
Table~\ref{Tab:ChemistrySpecies} lists the tracked species which are advected through the atmosphere, impact the radiative transfer calculation \citep[alongside the other, constant abundance, background atmospheric gases,][]{Boutle2017} and take part in the various chemical and photochemical reactions. The majority of the molecular opacities were sourced from the high-resolution transmission molecular absorption database \citep[HITRAN, ][]{Gordon2022}. Other sources include \citet{JPL2015} hereafter referred to as JPL2015, the MPI-Mainz UV/VIS Spectral Atlas \citep{Keller-Rudek2013}, and data from the South--West Research Institute \citep[SWRI,][]{Huebner2015}.

Our underlying chemistry framework \citep{Drummond2016} has previously been coupled to the dynamics and radiative transfer, and tested, at various levels of sophistication for performance such as conservation \citep[see][and Zamyatina et al., submitted]{Drummond2020}.

\begin{table*}
    \centering
    \caption{The species tracked in our model and involved in the ozone chemistry, and the source(s) for their opacity if the species is involved with the radiative transfer.}
    \label{Tab:ChemistrySpecies}
    \begin{tabular}{ccc}
        Species & Formula   &  Opacity data source         \\
        \hline \hline
        \multirow{2}{*}{Molecular oxygen}& \multirow{2}{*}{\ce{O2}} & HITRAN, recommended sources  for wavelengths \\
                                            &                       &  below 294\,nm in  \citet{JPL2015}  \\
        Ozone                               & \ce{O3}               & HITRAN, JPL2015, SWRI\\
        Molecular nitrogen                  & \ce{N2}               & HITRAN, \citet{Fennelly1992,Henke1993} \\
        Carbon dioxide                      & \ce{CO2}              & HITRAN, MPI-Mainz UV/VIS Spectral Atlas, SWRI \\
        Atomic oxygen (ground state)        & \ce{O(^3P)} & -- \\
        Atomic oxygen (first excited state) & \ce{O(^1D)} & -- \\ \hline
        \multirow{2}{*}{Water}              & \multirow{2}{*}{\ce{H2O}}  & HITRAN, MPI-Mainz UV/VIS Spectral Atlas,  \\
        & & SWRI\\
        Hydroxyl radical    & \ce{OH}                    & --\\
        Hydroperoxyl radical & \ce{HO2}                   & JPL2015, SWRI\\
        Molecular hydrogen   & \ce{H2}                    & --\\
        Atomic hydrogen      & \ce{H}                     & --\\
        Hydrogen peroxide    & \ce{H2O2}                  & JPL2015, SWRI\\ \hline
        Nitric oxide         & \ce{NO}                    & --\\
        Nitrogen dioxide     & \ce{NO2}                   & MPI-Mainz UV/VIS Spectral Atlas, SWRI\\
        Nitrate radical      & \ce{NO3}                   & MPI-Mainz UV/VIS Spectral Atlas, SWRI\\
        Dinitrogen pentoxide & \ce{N2O5}                  & MPI-Mainz UV/VIS Spectral Atlas, SWRI\\
        Peroxynitric acid    & \ce{HO2NO2}                & JPL2015, SWRI\\
        Nitrous acid         & \ce{HONO}                  & JPL2015, SWRI \\
        Nitric acid          & \ce{HNO3}                  & JPL2015, SWRI\\
        Nitrous oxide        & \ce{N2O}                   & HITRAN, MPI-Mainz UV/VIS Spectral Atlas  \\ 
        Atomic nitrogen (ground state)     & \ce{N(^4S)}  & --\\
        Atomic nitrogen (excited state)    & \ce{N(^2D)}  & --
    \end{tabular}
\end{table*}

\subsection{Reactions}
\label{appsubsec:reactions}
In this work we include several different forms of reactions which we class as bimolecular, termolecular and photolysis detailed in Sections~\ref{appsubsec:bimole}, \ref{appsubsec:termole} and \ref{appsubsec:photolysis}, respectively. Additionally, we  separately detail those reactions caused by SEPs in Section~\ref{appsubsec:sep_reactions}.

The parameters for the reactions are obtained from the following sources: \citet{JPL2015} (JPL2015), \citet{JPL2019} hereafter referred to as JPL2019, \citet{IUPAC} hereafter referred to as IUPAC, and for reactions involving atomic nitrogen \citet{Herron1999}.

\subsection{Bimolecular Reactions}
\label{appsubsec:bimole}
Bimolecular reactions are chemical reactions containing two reactants. The reaction rate ($k_f$, molecule\,cm$^{-3}$\,s$^{-1}$) of a bimolecular chemical reaction (with reactants \ce{A} and \ce{B}) is calculated as
\begin{equation}
    k_f = k [\ce{A}][\ce{B}],
\end{equation}
where $k$ is the reaction rate coefficient (cm$^{3}$\,molecule$^{-1}$\,s$^{-1}$) and [\ce{A}] is the number density of species \ce{A}, and [\ce{B}] the number density of species \ce{B} (both expressed in molecule\,cm$^{-3}$). The reaction rate coefficient is calculated using the modified Arrhenius equation
\begin{equation}
    k = A (T/300)^\alpha \exp(-E_a/RT),
\end{equation}
where $A$ is a pre--exponential factor (cm$^3$ molecule$^{-1}$ s$^{-1}$), $\alpha$ is a parameter which controls temperature dependence, $E_a$ is the activation energy of the reaction (J\,mol$^{-1}$), $R$ is the universal gas constant (8.3144 J\,K$^{-1}$\,mol$^{-1}$), and $T$ is the temperature. The parameters for every bimolecular reaction included in our model are included in Table~\ref{Tab:Bimolecular}.

\begin{table*}
    \centering
    \caption{The bimolecular reactions included in the chemical network, and their coefficients. \\ 
    Notes: \\ 1. In the presence of water there is an extra corrective factor of $1 + 1.4 \times10^{-21} [\ce{H2O}] \exp(2200/T)$. \\ 2. Integrated rate constant for both association and dissociation. \\ 3. We only use $k_1$ from the expanded rate coefficient to account for temperature dependence.} 
    \label{Tab:Bimolecular}
    \begin{tabular}{llllll}
        Reaction & A (cm$^3$ molecule$^{-1}$ s$^{-1}$)  & $\alpha$  & $E_a$/R (K) & T range (K) & Source \\ \hline  \hline
        \ce{O(^3P) + O3 -> O2 + O2} & $8.00\times10^{-12}$ & 0 & 2060 & 220-409 & JPL2019 \\ 
        \ce{O(^1D) + O2 -> O(^3P) + O2} & $3.3\times10^{-11}$ & 0 & -55 & 104-424 & JPL2019 \\
        \ce{O(^1D) + N2 -> O(^3P) + N2} & $2.15\times10^{-11}$ & 0 & -110 & 103-673 & JPL2019 \\
        \ce{O(^1D) + CO2 -> O(^3P) + CO2} & $7.5\times10^{-12}$ & 0 & -115 & 195-370 & JPL2019 \\
        \ce{O(^1D) + O3 -> O2 + O2} & $1.20\times10^{-10}$ & 0 & 0.0 & 103-393 & JPL2019 \\ 
        \ce{O(^1D) + O3 -> O2 + O(^3P) + O(^3P)} & $1.20\times10^{-10}$ & 0 & 0.0 & 103-393 & JPL2019 \\ 
        \hline
        \ce{O(^1D) + H2O -> OH + OH} & $1.63\times10^{-10}$ & 0 & -60 & 217-453 & JPL2019 \\
        \ce{O(^1D) + H2 -> OH +   H} & $1.20\times10^{-10}$ & 0 & 0 & 204-4210  & JPL2019 \\
        \ce{O(^3P) + OH -> O2 +   H} & $1.8\times10^{-11}$ & 0 & -180 & 136-515 & JPL2019 \\
        \ce{O(^3P) + HO2 -> O2 + OH} & $3.0\times10^{-11}$ & 0 & -200 & 229-391 & JPL2019 \\ 
        \ce{O(^3P) + H2O2 -> HO2 + OH} & $1.40\times10^{-12}$ & 0 & 2000 & 283-386 & JPL2019 \\
        \ce{H + O3 -> OH + O2} & $1.40\times10^{-10}$ & 0 & 470 & 196-424 & JPL2019 \\
        \ce{H + HO2 -> OH + OH} & $7.2\times10^{-11}$ & 0 & 0 & 245-300 & JPL2019 \\
        \ce{H + HO2 -> O(^3P) + H2O} & $1.6\times10^{-12}$ & 0 & 0 & 245-300 & JPL2019 \\
        \ce{H + HO2 -> H2 + O2} & $6.9\times10^{-12}$ & 0 & 0 & 245-300 & JPL2019 \\
        \ce{OH + O3 -> HO2 + O2} & $1.7\times10^{-12}$ & 0 & 940 & 190-357 & JPL2019 \\
        \ce{OH + H2 -> H2O + H} & $2.8\times10^{-12}$ & 0 & 1800 & 200-1050 & JPL2019 \\
        \ce{OH + OH -> H2O + O(^3P)} & $1.8\times10^{-12}$ & 0 & 0 & 233-580 & JPL2019 \\
        \ce{OH + HO2 -> H2O + O2} & $4.8\times10^{-11}$ & 0 & -250 & 252-420 & JPL2019 \\
        \ce{OH + H2O2 -> H2O + HO2} & $1.8\times10^{-12}$ & 0 & 0 & 200-300 & JPL2019 \\
        \ce{HO2 + O3 -> OH + O2 + O2} & $1.0\times10^{-14}$ & 0 & 490 & 197-413 & JPL2019 \\
        \ce{HO2 + HO2 -> H2O2 + O2} & $3.0\times10^{-13}$, $^1$ & 0 & -460 & 222-1120 & JPL2019 \\
        \hline
        \ce{O(^1D) + N2O -> N2 + O2} & $4.641\times10^{-11}$ & 0 & -20 & 195-719 & JPL2019 \\
        \ce{O(^1D) + N2O -> NO + NO} & $7.259\times10^{-11}$ & 0 & -20 & 195-719 & JPL2019 \\
        \ce{O(^3P) + NO2 -> NO + O2} & $5.1\times10^{-12}$, $^2$ & 0 & -210 & 199-2300 & JPL2015 \\
        \ce{O(^3P) + NO3 -> NO2 + O2} & $1.3\times10^{-11}$ & 0 & 0 & 298-329 & JPL2019 \\
        \ce{H + NO2 -> OH + NO} & $1.35\times10^{-10}$ & 0 & 0 & 195-2000 & JPL2019 \\
        \ce{OH + NO3 -> HO2 + NO2} & $2.0\times10^{-11}$ & 0 & 0 & 298 & JPL2019 \\
        \ce{OH + HONO -> H2O + NO2} & $3.0\times10^{-12}$ & 0 & -250 & 276-1400 & JPL2019 \\
        \ce{OH + HNO3 -> H2O + NO3} & $2.4\times10^{-14}$ & 0 & -460 & -- & IUPAC$^3$ \\
        \ce{OH + HO2NO2 -> H2O + NO2 + O2} & $4.5\times10^{-13}$ & 0 & -610 & 218-335 & JPL2019 \\
        \ce{HO2 + NO -> NO2 + OH} & $3.44\times10^{-12}$ & 0 & -260 & 182-1270 & JPL2019 \\
        \ce{HO2 + NO3 -> OH + NO2 + O2} & $3.5\times10^{-12}$ & 0 & 0 & 263-338 & JPL2019 \\
        \ce{N(^4S) + O2 -> NO + O(^3P)} & $3.3\times10^{-12}$ & 0 & 3150 & 280-1220 & JPL2019 \\
        \ce{N(^4S) + NO -> N2 + O(^3P)} & $2.1\times10^{-11}$ & 0 & -100 & 196-3660 & JPL2019 \\
        \ce{N(^4S) + NO2 -> N2O + O(^3P)} & $5.8\times10^{-12}$ & 0 & -220 & 223-700 & JPL2019 \\
        \ce{NO + O3 -> NO2 + O2} & $3.0\times10^{-12}$ & 0 & 1500 & 195-443 & JPL2019 \\
        \ce{NO + NO3 -> NO2 + NO2} & $1.7\times10^{-11}$ & 0 & -125 & 209-703 & JPL2019 \\
        \ce{NO2 + O3 -> NO3 + O2} & $1.2\times10^{-13}$ & 0 & 2450 & 231-362 & JPL2019 \\
        \ce{NO2 + NO3 -> NO + NO2 + O2} & $4.35\times10^{-14}$ & 0 & 1335 & 236-538 & JPL2019 \\
        \ce{NO3 + NO3 -> NO2 + NO2 + O2} & $8.5\times10^{-13}$ & 0 & 2450 & 298-1100 & JPL2019 \\
        \ce{N2O5 + H2O -> HNO3 + HNO3} & $2\times10^{-21}$ & 0 & 0 & 290-298 & JPL2019 \\
        \ce{N(^2D) + O(^3P) -> N(^4S) + O(^3P)} & $3.3\times10^{-12}$ & 0 & 260 & 300-400 & \citet{Herron1999} \\
        \ce{N(^2D) + O2 -> NO + O(^3P)} & $9.7\times10^{-12}$ & 0 & 185 & 200-500 & \citet{Herron1999} \\
        \ce{N(^2D) + N2O -> N2 + NO} & $1.5\times10^{-11}$ & 0 & 570 & 200-400 & \citet{Herron1999} \\
        \ce{N(^2D) + N2 -> N(^4S) + N2} & $1.7\times10^{-14}$ & 0 & 0 & 298 & \citet{Herron1999}
    \end{tabular}
\end{table*}

\subsection{Termolecular Reactions}
\label{appsubsec:termole}
A termolecular reaction is a reaction which involves three reactants. In this work, our termolecular reactions involve two main reactants and a third molecule (\ce{M}) which symbolises a range of possible third-body molecules. The third-body facilitates the reaction and stabilises the products.  The reaction rate ($k_f$) for a termolecular reaction is 
\begin{equation}
    k_f = k [\ce{A}][\ce{B}][\ce{M}],
\end{equation}
where $k$ is the reaction rate coefficient (cm$^{6}$\,molecule$^{-2}$\,s$^{-1}$) and [\ce{A}] is the number density of species \ce{A}, [\ce{B}] is the number density of species \ce{B}, and [\ce{M}] is the combined number density of all possible third-body molecules \ce{M}  (all expressed in molecule\,cm$^{-3}$). As such, the reaction rate coefficients of termolecular reactions are generally dependent on pressure. The low-pressure coefficient $k_0$ and the high-pressure coefficient $k_\infty$ are defined as,
\begin{equation}
k_0 = k_1 (T/300)^{\alpha_1}\exp(-\beta_1/T) \\ 
\end{equation}
and
\begin{equation}
k_\infty = k_2 (T/300)^{\alpha_2}\exp(-\beta_2/T),
\end{equation}
respectively. We determine the overall rate coefficient $k$ (cm$^6$ s$^{-1}$) using 
\begin{equation}
    k = k_0 (\frac{1}{1+P_r})F,
\end{equation}
where $P_r$ is the reduced pressure calculated using
\begin{equation}
    P_r = \frac{k_0 [M]}{k_\infty},
\end{equation}
where $[M]$ is the number density of the third-body molecule. $F$ is a broadening factor determined by 
\begin{equation}
    F = F_c^{1/\left(1+\left(\frac{\log_{10}(P_r) +c}{N-d(\log_{10}(P_r)+c)}\right)^2\right)},
\end{equation}
where $c = -0.4 -0.67\log_{10}(F_c)$, $N = 0.75 -1.27\log_{10}(F_c)$ and $d = 0.14$ and $F_c$ is calculated using
\begin{equation}
    F_c = (1-a) \exp(-T/T^{*}) + a \exp(-T/T^{**}) + \exp(-T^{***}/T),
\end{equation}
where $a$, $T^{*}$, $T^{**}$, and $T^{***}$ are parameters from the Troe formalism \citep{Troe1983}. In the case where all the Troe parameters are 0, $F= 1$, which is the Lindemann formalism \citep{Lindemann1922}. 

Tables~\ref{Tab:Termolecular} and \ref{Tab:TermolecularDecomp} present the complete list of all the termolecular reactions included in our model, and the relevant parameters. For Table~\ref{Tab:Termolecular}, some reactions are adequately described by the low pressure limit only therefore $k=k_0$ for these reactions and only the parameters for this value are listed. Decomposition reactions, where one reactant decomposes into two products, require significantly different values of the parameters so we present these separately in Table~\ref{Tab:TermolecularDecomp}.


\begin{table*}
    \centering
    \caption{The termolecular reactions included in the chemical network, and their coefficients (see Table~\ref{Tab:TermolecularDecomp} for decomposition reactions). \\ 
    Note: 1. In the presence of water there is an extra corrective factor of $1 + 1.4 \times10^{-21} [\ce{H2O}] \exp(2200/T)$.}
    \label{Tab:Termolecular}
    \begin{tabular}{llllllll} 
        Reaction & $k_1$ (cm$^{6}$\,molecule$^{-2}$\,s$^{-1}$) & $\alpha_1$ & $\beta_1$ (K) & $k_2$ (cm$^{3}$\,molecule$^{-1}$\,s$^{-1}$)  & $\alpha_2$ & $\beta_2$ (K) & Source \\ \hline  \hline
        \ce{O(^3P) + O2 + M -> O3 + M} & $6.1\times10^{-34}$& -2.4 & 0 & -- & -- & -- & JPL2019  \\ \hline
        \ce{HO2 + HO2 + M -> H2O2 + O2 + M} & $2.1\times10^{-33}$, $^1$ & 0 & -920 & -- & -- & -- & JPL2019\\
        \ce{H + O2 + M -> HO2 + M} & $5.3\times10^{-32}$ & -1.8 & 0 & $9.5\times10^{-11}$ & 0.4 & 0 & JPL2019 \\
        \ce{OH + OH + M -> H2O2 + M} & $6.9\times10^{-31}$ & -1 & 0 & $2.6\times10^{-11}$ & 0 & 0 & JPL2019 \\ \hline
        \ce{O(^1D) + N2 + M -> N2O + M} & $2.8\times10^{-36}$& -0.9 & 0 & -- & -- & -- & JPL2019 \\
        \ce{O(^3P) + NO + M -> NO2 + M} & $9.1\times10^{-32}$ & -1.5 & 0 & $3.0\times10^{-11}$ & 0 & 0 & JPL2019 \\
        \ce{O(^3P) + NO2 + M -> NO3 + M} & $3.4\times10^{-31}$ & -1.6 & 0 & $2.3\times10^{-11}$ & -0.2 & 0 & JPL2019 \\
        \ce{OH + NO + M -> HONO + M} & $7.1\times10^{-31}$ & -2.6 & 0 & $3.6\times10^{-11}$ & -0.1 & 0 & JPL2019 \\
        \ce{OH + NO2 + M -> HNO3 + M} & $1.8\times10^{-30}$ & -3 & 0 & $2.8\times10^{-11}$ & 0 & 0 & JPL2019 \\
        \ce{HO2 + NO2 + M -> HO2NO2 + M} & $1.9\times10^{-31}$ & -3.4 & 0 & $4\times10^{-12}$ & -0.3 & 0 & JPL2019 \\
        \ce{NO2 + NO3 + M -> N2O5 + M} & $2.4\times10^{-30}$ & -3 & 0 & $1.6\times10^{-12}$ & 0.1 & 0 & JPL2019
    \end{tabular}
\end{table*}

\begin{table*}
    \caption{The termolecular decomposition reactions included in the chemical network, and their coefficients.}
    \label{Tab:TermolecularDecomp}
    \centering
    \begin{tabular}{llllllll} 
        Reaction & $k_1$ (cm$^{3}$\,molecule$^{-1}$\,s$^{-1}$) & $\alpha_1$ & $\beta_1$ (K) & $k_2$ (s$^{-1}$)  & $\alpha_2$ & $\beta_2$ (K) & Source \\ \hline  \hline
        \ce{N2O5 + M -> NO2 + NO3 + M} & $1.3\times10^{-3}$ & -3.5 & 11000 & $9.7\times10^{14}$ & 0.1 & 11080 & IUPAC  \\
        \ce{HO2NO2 + M -> HO2 + NO2 + M} & $4.1\times10^{-5}$ & 0 & 10650 & $6.0\times10^{15}$ & 0 & 11170 & IUPAC
    \end{tabular}
\end{table*}

\subsection{Photolysis}
\label{appsubsec:photolysis}

The reaction rate ($k_f$) for a photolysis reaction is 
\begin{equation}
    k_f = J [\ce{A}],
\end{equation}
where $J$ is the reaction rate coefficient (molecule\,s$^{-1}$) and [\ce{A}] is the number density of species \ce{A} in molecule\,cm$^{-3}$. The reaction rate coefficients of photolysis reactions (or channels) are determined using
\begin{equation}
    J=\int_0^\infty Q(\lambda)\sigma(\lambda) F(\lambda) d\lambda,
\end{equation}
where $Q(\lambda)$ is the wavelength dependent quantum yield for each photolysis channel, $\sigma(\lambda)$ is the wavelength dependent cross section of the dissociating species, and $F(\lambda)$ is the actinic flux. These rates are calculated by the SOCRATES radiative transfer code \citep{Manners2022,Jackson2020}, and passed to our chemical solver \citep{Drummond2016}. Table~\ref{Tab:Photolysis} lists all the photolysis channels captured in our model, some of which are featured in the main paper text, but are repeated here for completeness. The threshold wavelength (corresponding to a photon with the minimum energy needed to dissociate the molecule), and the sources for the quantum yields are also included. Recommended quantum yields from JPL2019 \citep{JPL2019} were the primary source used in this work. When there were not recommended quantum yields, we assumed a quantum yield of 1 for all wavelengths. 

\begin{table*}
    \centering
    \caption{The list of photolysis reactions (channels) used in the chemical network, and the threshold wavelength for each reaction.}
    \label{Tab:Photolysis}
    \begin{tabular}{lll}
        Photolysis                            & Threshold wavelength (nm) & Quantum yield source\\ \hline  \hline
        \ce{O2 +     h{\nu} -> O(^3P) + O(^3P)}   & 242.3 & JPL2019\\
        \ce{O2 +     h{\nu} -> O(^3P) + O(^1D)}   & 175   & JPL2019\\
        \ce{O3 +     h{\nu} -> O(^3P) + O2}       & 1180  & JPL2019\\
        \ce{O3 +     h{\nu} -> O(^1D) + O2}       & 411   & JPL2019\\ \hline
        \ce{HO2 +    h{\nu} -> OH + O(^3P)}       & 438   & Assumed to be 1\\
        \ce{HO2 +    h{\nu} -> OH + O(^1D)}       & 259   & Assumed to be 1\\
        \ce{H2O +    h{\nu} -> H + OH}            & 242   & JPL2019\\
        \ce{H2O +    h{\nu} -> H2 + O(^1D)}       & 175   & JPL2019\\
        \ce{H2O +    h{\nu} -> H + H + O(^1D)}    & 175   & JPL2019\\
        \ce{H2O2 +   h{\nu} -> OH + OH}           & 557   & JPL2019\\
        \ce{H2O2 +   h{\nu} -> H + HO2}           & 557   & JPL2019\\ \hline
        \ce{NO2 +    h{\nu} -> NO + O(^3P)}       & 422   & JPL2019\\ 
        \ce{NO3 +    h{\nu} -> NO2 + O(^3P)}      & 7320  & JPL2019\\ 
        \ce{NO3 +    h{\nu} -> NO + O2}           & 640   & JPL2019\\
        \ce{N2O +    h{\nu} -> N2 + O(^1D)}       & 336   & JPL2019\\
        \ce{N2O5 +   h{\nu} -> NO3 + NO2}         & 1255  & JPL2019\\
        \ce{N2O5 +   h{\nu} -> NO3 + NO + O(^3P)} & 298   & JPL2019\\
        \ce{HONO +   h{\nu} -> OH + NO}           & 579   & Assumed to be 1\\
        \ce{HNO3 +   h{\nu} -> NO2 + OH}          & 581   & Assumed to be 1\\
        \ce{HO2NO2 + h{\nu} -> HO2 + NO2}         & 1207  & JPL2019\\
        \ce{HO2NO2 + h{\nu} -> OH + NO3}          & 726   & JPL2019
    \end{tabular}
\end{table*}

\subsection{Stellar Proton Forcing}
\label{appsubsec:sep_reactions}
The impact of SEPs is discussed in Section~\ref{subsec:protons}, and the complete list of reactions we include in our model is given in Table~\ref{Tab:StellarProtons}. The reactions caused by SEPs are described in the main paper, but repeated here for completeness.

\begin{table*}
    \centering
    \caption{The list of reactions caused by stellar proton (or stellar energetic particles, SEPs) forcing used in the chemical network and the total amount of molecules produced per ion pair for each reaction.}
    \label{Tab:StellarProtons}
    \begin{tabular}{ll}
        Reaction                   & Production efficiency \\ \hline  \hline
        \ce{H2O -> H + OH}         &  2    \\
        \ce{N2 -> N(^4S) + N(^4S)} &  0.55    \\
        \ce{N2 -> N(^2D) + N(^2D)} &  0.7    
    \end{tabular}
\end{table*}

\clearpage

\section{Planetary habitability and observability including the contribution of nitric acid}
\label{appsec:withHNO3}
This appendix section shows versions of Figures~\ref{Fig:UV}-\ref{Fig:Transmission} from Sections~\ref{Sec:PlanetHab} and \ref{Sec:PlanetObs} with the contribution of nitric acid (\ce{HNO3}) included in the calculations. 

\subsection{Habitability}

\begin{figure*}
    \centering
    \includegraphics[width=\textwidth]{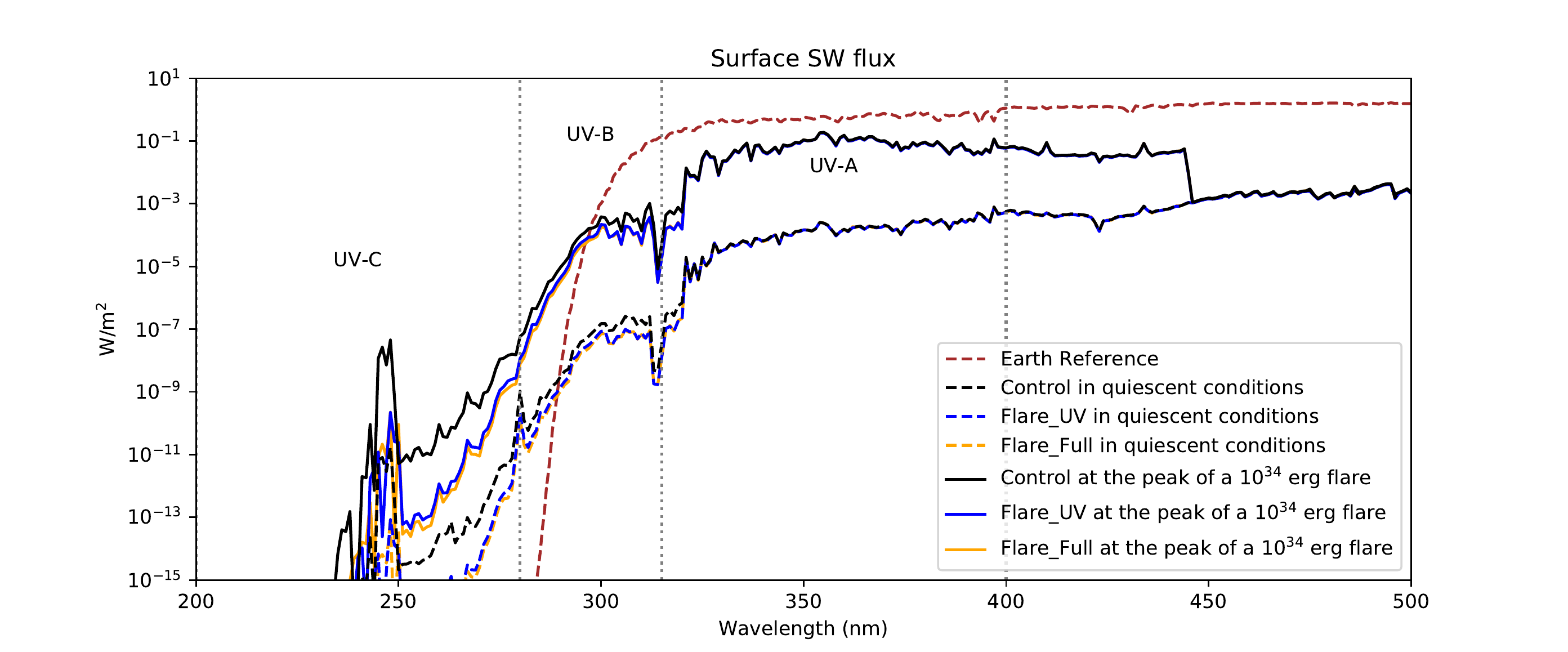}
    \caption{The average day-side surface UV environment under quiescent conditions and during the peak of a $10^{34}$\,erg flare from the end of the spin-up simulation Quiet\_Full (black), a run containing only stellar flares (blue), and a simulation containing both stellar flares and CMEs (orange), using the atmospheric configurations from the end of their respective simulations, including the contribution of \ce{HNO3}. The changes in atmospheric composition due to stellar flares have caused additional screening of the surface from UV radiation. The inclusion of \ce{HNO3} has caused a significant reduction in the UV--B and UV--C fluxes.}
    \label{Fig:UVwithHNO3}
\end{figure*}

The large amount of \ce{HNO3} in our simulations causes significant changes in the surface UV radiation environment. Figure~\ref{Fig:UVwithHNO3} shows significantly smaller UV--B and UV--C fluxes which reach the planet's surface. The amount of UV--A is unchanged. The reduction in UV--C is quite important due to the potential harm to life.

\begin{figure*}
    \centering
    \includegraphics[width=\textwidth]{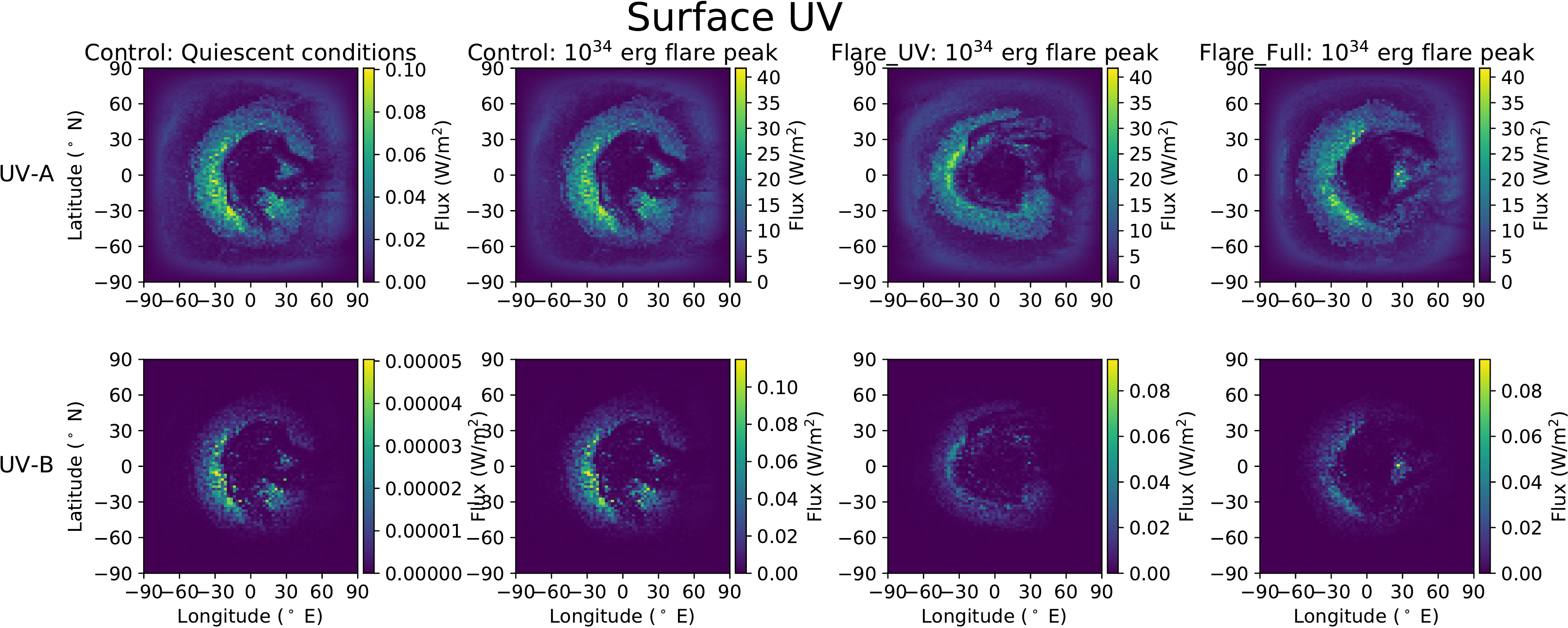} 
    \caption{The surface UV radiation environment for the Control simulation under quiescent conditions, as well as Control, Flare\_UV, and Flare\_Full if they were to be subject to the peak of a $10^{34}$\,erg flare, including the contribution of \ce{HNO3} to radiative absorption. As seen previously in Figure~\ref{Fig:UVwithHNO3}, the changes in atmospheric composition due to flares have not significantly changed the UV--A flux, but have resulted in a a minor reduction in UV--B.}
    \label{Fig:QuiUVwithHNO3}
\end{figure*}

Figure~\ref{Fig:QuiUVwithHNO3} shows the spatial distribution of the UV--A and UV--B fluxes. The flux levels of UV--C are very low in this case, where \ce{HNO3} is not removed from the atmosphere. We observe that the additional screening caused by the changes in atmospheric composition now result in a small reduction in UV--B. 

\begin{figure*}
    \centering
    \includegraphics[width=\textwidth]{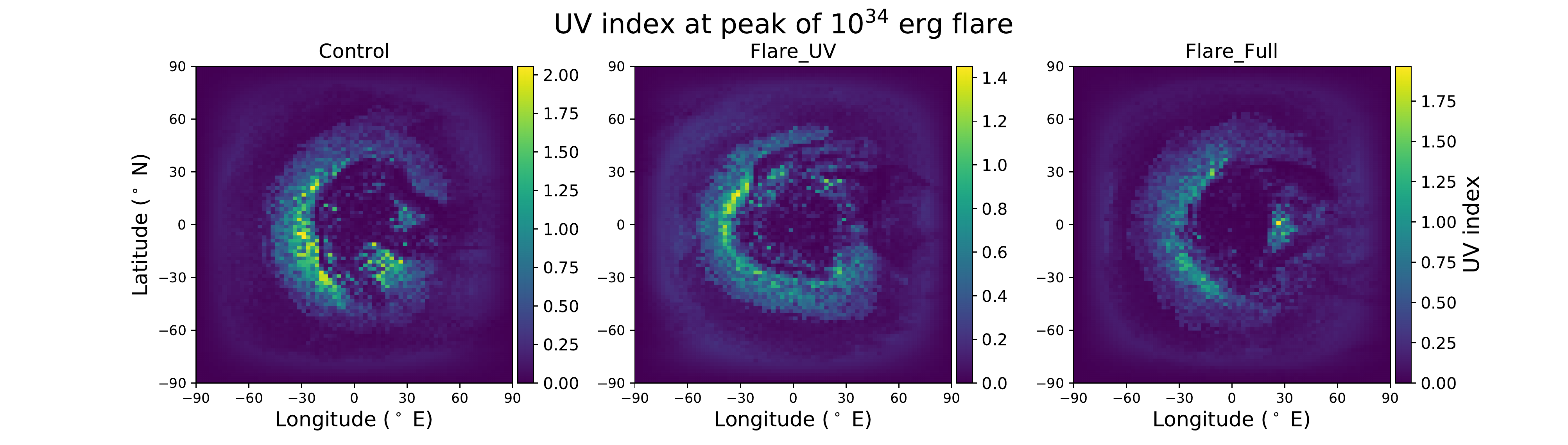}
    \caption{Maps of the UV index for Control, Flare\_UV, and Flare\_Full if they were to be subject to the peak of a $10^{34}$\,erg flare, including the contribution of \ce{HNO3}.}
    \label{Fig:UVIndexwithHNO3}
\end{figure*}

Figure~\ref{Fig:UVIndexwithHNO3} shows the UV index at the peak of a 10$^{34}$\,erg flare. The additional screening due to \ce{HNO3}'s contribution causes the UV index to drop substantially. It is now very mild, and is not significantly reduced by the changes in atmospheric composition caused by flares.

\subsection{Observability}

\begin{figure*}
    \centering
    \includegraphics[width=\textwidth]{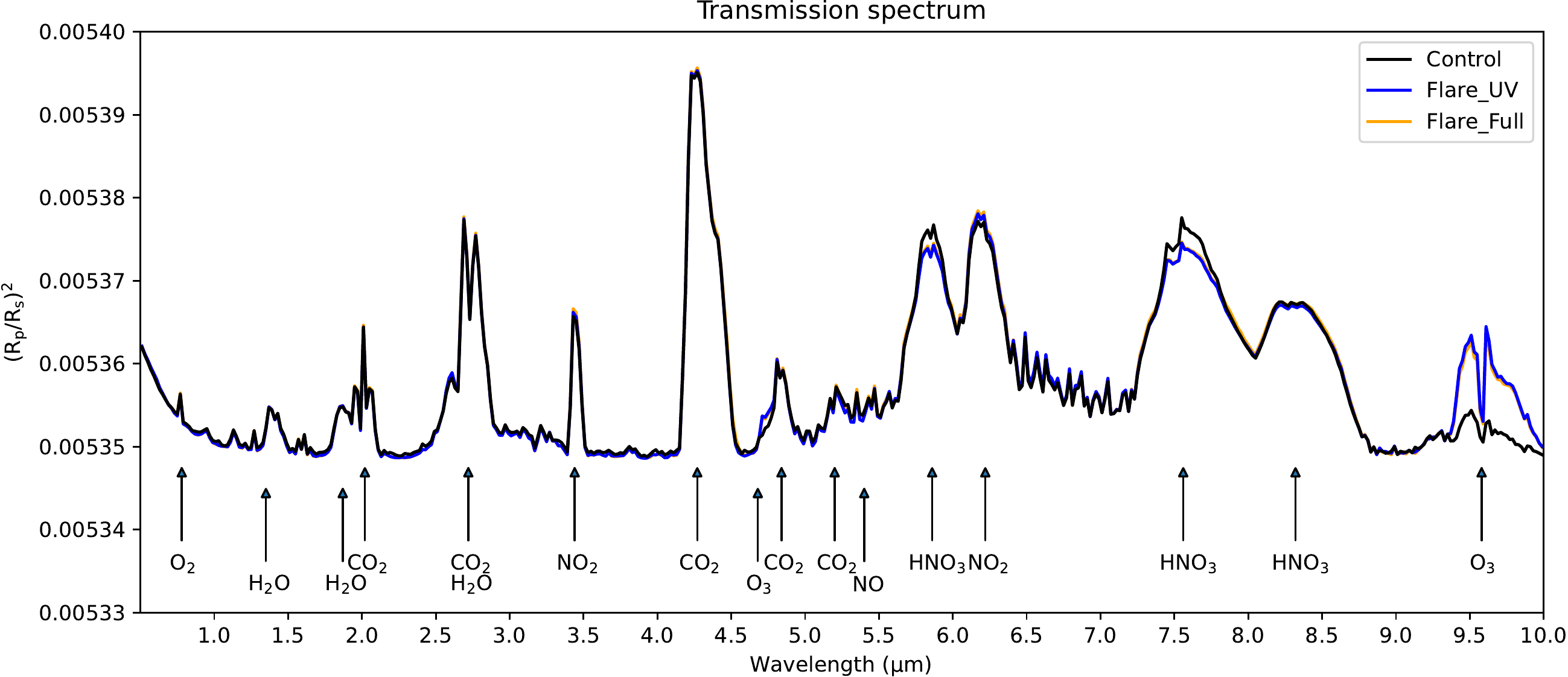}
    \caption{The transmission spectra for the simulated planets for 500\,nm-10\,$\mu$m at the end of their respective simulations including the contribution of \ce{HNO3}. The colors refer to the same simulations as Figure~\ref{Fig:UV}.}
    \label{Fig:TransmissionwithHNO3}
\end{figure*}

Figure~\ref{Fig:TransmissionwithHNO3} shows the transmission spectra With the contribution of \ce{HNO3}. We see that \ce{HNO3} has absorption features at 5.8, 7.5, and 8.3\,$\mu$m, but is otherwise quite similar to Figure~\ref{Fig:Transmission}.

\bsp	
\label{lastpage}
\end{document}